\documentclass[%
 reprint,
 amsmath,amssymb,
 aps,
]{revtex4-2}

\usepackage{graphicx}
\usepackage{dcolumn}
\usepackage{bm}
\usepackage{soul}
\usepackage{xcolor}

\newcommand{\be}{\begin{equation}}
\newcommand{\ee}{\end{equation}}
\newcommand{\bea}{\begin{eqnarray}}
\newcommand{\eea}{\end{eqnarray}}

\begin{document}

\preprint{APS/123-QED}

\title{Field-dependent magnetic relaxation times of magnetic nanoparticle systems: \\analytic approximations supported by numerical simulations}

\author{Jonathon C. Davidson}
 \affiliation{Biofrontiers Center, University of Colorado Colorado Springs, Colorado Springs, Colorado, 80918 USA.}

 \author{Nicholas R. Anderson}
 \affiliation{Biofrontiers Center, University of Colorado Colorado Springs, Colorado Springs, Colorado, 80918 USA.}

\author{Karen L. Livesey}
\affiliation{
School of Information and Physical Sciences, The University of Newcastle, Callaghan, NSW, 2308 Australia
}
\affiliation{Biofrontiers Center, University of Colorado Colorado Springs, Colorado Springs, Colorado, 80918 USA.}

\date{\today}

\begin{abstract}

Many estimates for the magnetic relaxation time of magnetic nanoparticle systems neglect the effect of the applied field strength. This is despite many applications of magnetic nanoparticles involving relaxation dynamics under the influence of applied fields. Here, an analytic approximation for the field-dependent Brownian relaxation time of single-domain, spherical magnetic nanoparticles in an external applied field is developed mathematically. This expression is validated by comparison with existing empirically-derived expressions and by comparison to particle-level simulations that allow particle rotations. Our approximation works particularly well for larger particles. We then use the developed expression to analytically calculate the total magnetic relaxation time when both Brownian and N\'eel relaxation mechanisms are at play. Again, we show that the results match those found using particle-level simulations, this time with both particle rotations and internal magnetization dynamics allowed. However, for some particle parameters and for large field strengths, our simulations reveal that the Brownian and N\'eel relaxation mechanisms are decoupled and it is not appropriate to combine these to calculate a total relaxation time.

\end{abstract}

\maketitle


\section{\label{sec:level1}Introduction}

Magnetic nanoparticles (MNPs) have gained significant attention in various research fields, ranging from biomedicine to energy and environmental applications.~\cite{shasha2021nonequilibrium,tang2013magnetic} Understanding the fundamental properties of MNPs, such as their magnetic relaxation times, is crucial for optimizing their performance and unlocking their potential in these applications.~\cite{deatsch2014heating,torres2019relevance,rajan2020assessing,giri2005preparation,croft2012relaxation}

Magnetic relaxation time refers to the characteristic time required for {an ensemble of }MNPs to reach their equilibrium magnetic state. The relaxation mechanisms include N\'eel relaxation and Brownian relaxation. N\'eel relaxation occurs when the net magnetic moment of an individual nanoparticle reorients internally (often overcoming an anisotropy energy barrier), while Brownian relaxation arises from the rotation of the entire nanoparticle to achieve {dynamic equilibrium.}~\cite{deatsch2014heating} Which mechanism dominates the dynamics {as the ensemble moves toward} equilibrium depends on numerous material and system parameters such as the size and composition of the nanoparticles and the viscosity of the liquid suspension. Generally speaking, larger nanoparticles are dominated by Brownian relaxation mechanism while smaller particles are controlled through N\'eel relaxation.~\cite{deissler2013brownian,rosensweig2002heating}

The understanding of magnetic relaxation times is essential due to its direct implications on the performance and functionality of MNPs in technological applications. One notable area where magnetic relaxation times play a pivotal role is in magnetic particle imaging (MPI). In MPI, MNPs are used as tracer particles and enable the capture of real time  and quantitative images from inside the human body. The relaxation times of MNPs are one factor in determining the image contrast of the specific particle used.~\cite{croft2012relaxation,wu2019review} 

Furthermore, magnetic relaxation times affect the effectiveness of magnetic hyperthermia, a promising cancer treatment. By subjecting MNPs to an alternating magnetic field, localized heating can be induced, leading to targeted tumor ablation.~\cite{bekovic2023magnetic} The main parameter used to measure the conversion of magnetic energy into thermal energy is the specific absorption rate (SAR). The mechanisms for heating and the SAR measurement depend on material parameters such as the Brownian and N\'eel relaxation time of the particles.~\cite{rosensweig2002heating} This means the optimization of magnetic relaxation times allows for precise control over the heating efficiency and ensures selective destruction of cancerous cells, minimizing damage to healthy tissues.~\cite{deatsch2014heating,rajan2020assessing,fortin2008intracellular}

However, the dependence of the magnetic relaxation time on the applied field strength is often not included in analyses of MNPs~\cite{bogren2015classification} and can {result} in orders of magnitude difference for the relaxation time.~\cite{deissler2013brownian}
In our previous work, we obtained an accurate expression for the \emph{N\'eel} relaxation time~\cite{chalifour2021magnetic} of randomly-oriented MNPs in an applied field. Here, a simple but novel approach to estimate the \emph{field-dependent Brownian} magnetic relaxation time is presented, {when an applied field is suddenly switched on}. In order to validate the found expression we use particle-level dynamics simulations. The simulations provide a comprehensive understanding of the behavior of MNPs at the individual particle scale. This enables us to calculate the characteristic relaxation time of an ensemble of nanoparticles in various conditions.

Previous studies have investigated Brownian relaxation times and produced field-dependant expressions. These expressions are either empirically fitted to numerical simulation data~\cite{yoshida2009simulation} or are analytic approximations~\cite{martsenyuk1974kinetics} {which have been verified through  additional experiments.}~\cite{dieckhoff2016magnetic} {Martsenyuk \emph{et al.}}~\cite{martsenyuk1974kinetics} {derived  perpendicular and parallel Brownian relaxation times, $\tau_{\perp}$ and $\tau_{||}$, corresponding to the characteristic times for the net moment components perpendicular and parallel to the applied field. This was done in two ways within the same article; first, by examining the smallest eigenvalue of the Fokker-Planck equation (which is reminiscent of Brown's work on N\'eel relaxation~}\cite{brown1963thermal}), {and second, by taking a thermal average of the microscopic equation of motion for a moment. It is $\tau_{\perp}$ (written in the Appendix) that we can compare our own expression to, since we consider a random ensemble of particles which align their moments when a field is switched on.} 

{The experiments to confirm the expressions for $\tau_{\perp}$ and $\tau_{||}$~}\cite{dieckhoff2016magnetic}{ applied a large static magnetic field and then the ensemble's response to an additional, small probing field -- either perpendicular or parallel to the static field -- was measured. In this case, the probability density of magnetization angles changes very little when the probing field is added. We ask a different question here. If there is initially no applied field and the ensemble has a random configuration of magnetic moments, then how long does it take the ensemble to respond to an applied field turned on instantly?}

{We note that an article by Deissler \emph{et al.}}~\cite{deissler2014dependence} {also investigates how Brownian and N\'eel relaxation times (separately) vary with applied field strength. However, their results were numerical. Here, we present an analytic expression, which has the potential for wide adoption due to its ease of use.}

After examining the field-dependent Brownian relaxation time, we calculate the total magnetic relaxation time for a system of MNPs that are oriented at random and are placed in a magnetic field. Again, we compare the results of analytic approximations with those from full particle-level simulations and conclude that analytic approximations may provide great utility to those wishing to design materials with a specific magnetic relaxation time. The total relaxation time combines both N\'eel and Brownian relaxation mechanisms, as is appropriate for particles in a fluid. However, particle \emph{translations} are not yet considered in this work.

Additionally, while studying the total field-dependent relaxation time we found evidence of a phenomenon reported on in the literature.\cite{ota2016rotation} It is generally assumed that both N\'eel and Brownian relaxations happen at the same time.~\cite{rosensweig2002heating} However, recent work suggests this may not always be the case \cite{ota2016rotation} and after initial N\'eel relaxation occurs, slower Brownian relaxation can then dominate the dynamics of the particle, even for small nanoparticles. We find evidence for this in moderate fields of 200~Oe (20~mT). In this case, the typical analytic approach for combining N\'eel ($\tau_N$) and Brownian ($\tau_B$) relaxation times reciprocally to find the total relaxation time,~\cite{rosensweig2002heating} i.e.
\begin{equation}
    \frac{1}{\tau_{\textrm{total}}} = \frac{1}{\tau_{N}} + \frac{1}{\tau_{B}},
\end{equation}
is not appropriate, as we will demonstrate in this article. The fact that two different relaxation mechanisms are decoupled is important, for example, if one is interested in the timescales with which MNPs produce heat in hyperthermia.

In Section~\ref{meth} the analytic method to estimate the field-dependent Brownian relaxation time is described. The subsequent Section~\ref{Sim} details the equations of motion and  { material parameters used in the particle-level simulations that consider magnetization dynamics plus particle rotations.}~\cite{lemons1997paul}  Finally, Section~\ref{res} compares the developed expression and simulation results, and discusses the implications for estimating MNP relaxation time. MNP parameters appropriate for magnetite are used throughout the article.

\section{Analytic estimates for field-driven Brownian relaxation time}
\label{meth}

\begin{figure}
\centering
\includegraphics[width = 0.50\textwidth]{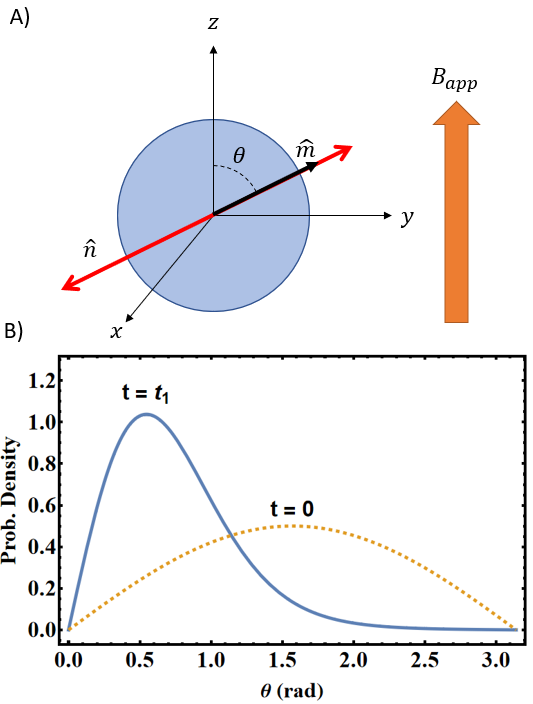}
\caption{(a) Coordinate system for a magnetic nanoparticle, with an applied field in the $z$ direction. The easy axis $\hat{n}$ (red double-arrow) and the magnetization unit vector $\hat{m}$ (black single-arrow) point in the same direction when calculating purely Brownian motion. They makes an angle of $\theta$ with the $z$ direction.  (b) The probability density of magnetization angles at time $t=0$ (dashed line, random orientations) and at long times (solid line, applied field $B=10$~mT causing partial alignment).}
\label{Coord}
\end{figure}

In order to obtain an analytic expression for the Brownian relaxation time in an applied magnetic field, a few basic assumptions are made. Each particle is a perfect sphere with a radius $r$, with uniaxial anisotropy axis in the direction denoted by unit vector $\hat{n}$, and with a magnetization direction denoted by unit vector $\hat{m}$. Each particle in a sample may have a different $\hat{n}$ and $\hat{m}$. For Brownian motion alone (ignoring N\'eel relaxation for now) it is assumed that $\hat{n}$ and $\hat{m}$ are pointing in the same direction within each particle. In other words, the magnetization in each particle is locked to the easy axis direction.~\cite{berkov2006langevin} The applied field direction is along $z$, as illustrated in  Figure~\ref{Coord}(a). The angle between the easy axis/magnetic moment and the applied field is $\theta$. If we use the properties of magnetite (Fe$_{3}$O$_{4}$), the particles have a single magnetic domain in the size regime investigated, namely $r=6-16$~nm.~\cite{kittel1949physical,reichel2017single} 

Obtaining an analytic approximation for the relaxation time of an ensemble of MNPs involves going beyond the single particle picture. Rather than calculating for each individual nanoparticle's thermal equations of motions, one typically examines the evolution of the probability density function $f(\theta,t)$ in time $t$. $f(\theta,t)$ is the probability density to measure a magnetic moment per unit angle $\theta$, at time $t$. The Fokker–Planck equation describes the time evolution of the distribution of magnetic moments.~\cite{brown1963thermal,chalifour2021magnetic} Figure~\ref{Coord}(b) shows two probability densities: at $t=0$ before any field is turned on (dashed line), and at $t=t_{1}$, a later time when equilibrium has been reached with a positive applied field (solid line). Note that without the field (dashed line) the probability density is centered on the equator, while the addition of the field shifts it towards smaller angles, with a peak whose width depends on the temperature. The characteristic time we are solving for describes how long it takes for the uniform density function at $t=0$ to move towards equilibrium at $t=t_{1}$. 

Following the same treatment as Debye with polar molecules rotating in an electric field~\cite{debye1929polar}, one can arrive at the equation of motion
\be
\gamma_r \frac{\partial f}{\partial t} = \frac{1}{\sin{\theta}}\frac{\partial}{\partial \theta}\left[ \sin{\theta}\left(k_{B}T\frac{\partial f}{\partial \theta} - M _{tor} f \right) \right],
\label{original}
\ee
where $\gamma_r =8\pi\eta r^{3}$ is a constant measuring the friction for a sphere, with $\eta$ the viscosity of the fluid and $r$ the particle radius.~\cite{debye1929polar,anderson2021simulating,wang2002molecular}
Note that here we are assuming that the hydrodynamic radius and the magnetic radius of the particles are \emph{the same}, but typically $\gamma_r =8\pi\eta R^{3}$, with hydrodynamic radius $R$ greater than the magnetic radius $r$ ($R>r$). This simplification can easily be relaxed and is a subject for future work. It is not limiting here where theory will be compared to simulations of MNPs. 

In Eq.~\eqref{original}, $k_{B}$ is the Boltzmann constant, $T$ is temperature, and $M_{tor}$ is the torque on the moment due to the field, namely
\be
M_{tor} = -\mu B \sin{\theta} = -\frac{\partial U}{\partial \theta} ,
\ee
with the negative indicating the moment will move in the direction of the field $B$. $U$ denotes the potential energy producing the torque due to the applied field $B$. Here, $\mu= M_{sat} V$ is the magnitude of magnetic particle moment with saturation magnetization $M_{sat}$ and volume $V$. Previously this differential equation has been solved in the case of $B=0$, arriving at the commonly cited Brownian relaxation time \cite{debye1929polar}
\be
\tau_{B} =\frac{\gamma_{r}}{2k_{B}T} = \frac{3 \eta V}{k_{B} T}.
\label{ogbrown}
\ee
However, we are interested in deriving a relaxation time when the particles are subjected to an applied field. 

With a change of variables, $x= \cos\theta$, Eq.~\eqref{original} becomes
\be
\gamma_r \frac{\partial f}{\partial t} = \frac{\partial}{\partial x} \left[(1-x^{2}) \left[ k_{B} T\frac{\partial f}{\partial x}+\mu B f \right] \right] .
\label{xvary}
\ee
In order to solve this differential equation analytically, we ignore the nonlinear $x^2$ term. The validity of this assumption is discussed here by considering the variable $x= \cos\theta$ and the initial distribution of the probability density $f(\theta,t=0)$.  When starting with a random distribution of magnetic moments, $f(\theta,t=0)$, the majority of particles have their moment near the equator ($\theta \sim \frac{\pi}{2}$) as shown in Figure~\ref{Coord}(b).~\cite{weisstein2002sphere} In turn, the $x^2$ term is small when $\theta \sim \frac{\pi}{2}$ and can be removed from Eq.~\eqref{xvary} in order to simplify. Note that when the distribution evolves due to the presence of the applied field (see Fig.~\ref{Coord}(b)) this assumption is no longer valid and the nonlinear components will play a role. We will compare the approximate solution for the relaxation time under this assumption to the relaxation time found from simulations in the results section, to ultimately check its validity.

Eq.~\eqref{xvary} when the nonlinear term is ignored becomes
\be
\gamma_{r} \frac{\partial f}{\partial t} = k_{B}T\frac{\partial^{2} f}{\partial x^{2}}+ \mu B \frac{\partial f}{\partial x}.
\ee
One can then transform the coordinate system and the corresponding derivatives by defining 
\bea
\tilde{x}=& x +\frac{\mu B}{\gamma_{r}} t & \longrightarrow \frac{\partial f}{\partial x} =\frac{\partial f}{\partial \tilde{x}}
\label{transform}
\\
\tilde{t}=& t~~~~~~~~~&  \longrightarrow \frac{\partial f}{\partial t} =\frac{\partial f}{\partial \tilde{x}}\left(\frac{\mu B}{\gamma_{r}}\right) + \frac{\partial f}{\partial\tilde{t}}
\nonumber .
\eea
This is equivalent to transforming the driven diffusion equation to the regular diffusion equation, via transformation to a moving coordinate system.~\cite{jensen1980solution}

Finally, the equation of motion for the MNP distribution function becomes
\be
\frac{\partial f}{\partial t} = \frac{k_B T}{\gamma_{r}}\frac{\partial^{2} f}{\partial \tilde{x}^{2}}.
\label{diffusion}
\ee
This is a one dimensional diffusion equation with two characteristic timescales. The first timescale comes from the thermal diffusion of the density of magnetic moments, described by Eq.~\eqref{diffusion}. Eq.~\eqref{diffusion} is, in fact, equivalent to Eq.~\eqref{original} with an applied field of $B=0$, which results in the standard Brownian relaxation time \cite{debye1929polar} for MNPs in zero field, given in Eq.~\eqref{ogbrown}. 

The second timescale comes from the driving of $f$ towards smaller angles due to the applied field and mathematically is determined by the ``velocity" of the coordinate transformation. Reading from Eq.~\eqref{transform}, the effective velocity is $\frac{\mu B}{\gamma_{r}}$, which has units of inverse time.
The time scale of the driven motion is therefore
\be
\tau_{\textrm{driven}} = \frac{\gamma_r}{\mu B}.
\ee

Since the fastest mechanism tends to dominate the Brownian relaxation time in a field $\tau_{BF}$, one can combine the diffusive and driven times reciprocally to get 
\be
\frac{1}{\tau_{BF}} = \frac{1}{\tau_{B}} + \frac{1}{\tau_{\textrm{driven}}},
\label{timecombine}
\ee
This is in the spirit of combining N\'eel and Brownian relaxation times to get the total relaxation time in zero field.~\cite{martsenyuk1974kinetics,rosensweig2002heating} 
This gives an approximate total \emph{Brownian} relaxation time \emph{in an applied field} of
\be
{\tau_{BF}} \sim \frac{\gamma_{r}}{2k_B T + \mu B}.
\label{Brownianfield}
\ee
Note that in the zero magnetic field ($B=0$), the relaxation time collapses to the original Brownian relaxation time. 

We now consider the case of strong magnetic fields, or relatively large-radius particles, given by $2k_{B} T << \mu B$. For magnetite at room temperature and a field strength of $B=10$~mT the competing thermal and magnetic energies are equal when the radius of the nanoparticles are $r \approx 8.5$ nm. So the larger radius limit corresponds to, say $r>10$~nm. Eq.~\eqref{Brownianfield} for the field-dependent Brownian relaxation time can then be written in the \emph{large radius limit} as
\be
{\tau_{BF~\textrm{large}}} \simeq\frac{\gamma_r }{\mu B} = \frac{8 \pi \eta r^{3} }{M_{sat} V B} = \frac{6 \eta  }{ M_{sat}  B}.
\label{BrownianfieldApprox}
\ee
Notice the volume term cancels out and Eq.~\eqref{BrownianfieldApprox} is independent of size. This is only strictly true when the hydrodynamic radius and the magnetic radius of the particle are the same, as we have considered here. The expression is a little more complicated when these radii are different. {Additionally,}  Eq.~\eqref{BrownianfieldApprox} {and the large field approximation for $\tau_{\perp}$ made in }Ref.~\cite{martsenyuk1974kinetics} {(see Eq.}~\eqref{martyLarge} {in the Appendix) collapse to the same function when the hydrodynamic radius and the magnetic radius are equal. This is a good check of our analytic result.}

The validity of  Eq.~\eqref{BrownianfieldApprox} will be examined in the results Sec.~\ref{res}, comparing it to other expressions in the literature, plus the results of simulations. {Existing analytic expressions for Brownian relaxation time in a field are provided in the Appendix.} In Sec.~\ref{res} we will also combine this developed field-dependent Brownian relaxation time with a field-depenedent N\'eel relaxation time~\cite{chalifour2021magnetic} to examine total relaxation times.

\section{Simulations}
\label{Sim}

A particle-level Langevin dynamics simulation is developed using a home-built Fortran code to investigate the dynamics of MNP moments.~\cite{lemons1997paul}  Unless otherwise stated, the simulations are done at 300~K and are comprised of 500 nanoparticles, each assumed to be perfect spheres with a radius of $r$, made of magnetite (Fe$_{3}$O$_{4}$) and suspended in {blood to mimic an environment suitable for magnetic hyperthermia treatment.} The density of the magnetite is assumed to be $\rho = 5255$~kg/m$^3$. The magnetic saturation and anisotropy constant used are $M_{sat} = 312$ kA/m and $K= 10$ kJ/m$^{3}$ respectively.~\cite{saville2014formation,mamiya2020estimation,fannin1994calculation,yoon2011determination} {While magnetite has a cubic anisotropy, here we model it with uniaxial anisotropy. Few articles have considered the effect of the cubic anisotropy on dynamic properties. An exception is}~\cite{zhao2020effects} {which showed through simulations that particles with cubic anisotropy have relaxation times that do not vary with the anisotropy constant. Here we focus on uniaxial anisotropy to compare to other calculations of the relaxation time, and future work can include extending our method to cubic anisotropy.} Note that magnetite nanoparticles with radius less than 40~nm~\cite{reichel2017single} are single domain and so a single macrospin $\hat{m}$ is appropriate to model each particle. The starting configuration of each particle's macrospin moment is randomly oriented, which gives a total magnetization of zero for the system at $t=0$. An applied field of 0, 10, or 20~mT is applied in the positive $z$-axis direction and the evolution of the net magnetization versus time is calculated. These calculations takes several hours on a computer cluster of 25 Intel(R) Xeon 2.2~GHz processors.

In order to predict the motion of the magnetic moments and easy axes, a set of coupled equations of motion given in Engelmann \emph{et al.} \cite{engelmann2019predicting} is used. The Landau-Lifshitz equation of motion governs the macrospin moment of each particle, indexed by $i$, and is given by
\be
\frac{d\hat{m}_{i}}{dt} = \frac{\mu_{0}\gamma}{1+\alpha} \Big( \vec{H}_{\textrm{eff},i} \times \hat{m}_{i} + \alpha \hat{m}_{i} \times (\vec{H}_{\textrm{eff},i} \times \hat{m}_{i}) \Big), 
\label{LL}
\ee
where $\hat{m}_{i}$ is a unit vector parallel to the moment of particle $i$, $\mu_{0}$ is the permeability of free space, $\gamma$ is the  gyromagnetic ratio of an electron as $1.76\times10^{11}$ rad$\cdot$Hz/T, $\alpha${=0.1} is the damping constant, and $\vec{H}_{\textrm{eff},i}$ is the effective field felt by particle $i$. This field is
\be
\vec{H}_{\textrm{eff},i} = \vec{H}_{app} + \frac{2KV}{\mu \mu_{0}}(\hat{m}_{i} \cdot \hat{n}_{i})\hat{n}_{i} + \vec{H}_{th},
\ee
with contributions from the applied field, the uniaxial anisotropy field, and stochastic thermal field respectively. $K$ is the anisotropy constant and $V$ is the volume of the MNP. $\hat{n}_{i}$ is the easy axis vector. Note that dipole-dipole interactions are ignored here and particles are non-interacting. Indeed, finding the magnetic relaxation time of interacting systems is an open problem which is important to address for real MNP systems.~\cite{dormann1997magnetic,ilg2020dynamics}

The motion of the easy axis vector $\hat{n}_{i}$ is governed by
\be
\frac{d\hat{n}_{i}}{dt} = \frac{\vec{\Theta}_{i}}{6 \eta V} \times \hat{n}_{i},
\label{rotate}
\ee
where $\vec{\Theta}_i$ is the total torque on the particle $i$ and $\eta$ is the liquid viscosity. {Here we use the viscosity of blood as $3.5$ mPa$\cdot$s}.~\cite{nader2019blood} The torque can be written as 
\be
\vec{\Theta}_{i} = -2K V( \hat{m}_{i} \cdot \hat{n}_{i})
(\hat{m}_{i} \times \hat{n}_{i}) + \vec{\Theta}_{th},
\ee
with the first term being the torque due to the separation of the magnetic moment from the easy axis, and the second term being a stochastic, thermal torque. 

Stochastic magnetic fields and mechanical torques model the role of temperature and collisions with the fluid. They are related to the magnetic damping $\alpha$ and the fluid viscosity $\eta$ via the fluctuation-dissipation theorem.~\cite{wang2002molecular} Namely, the correlation functions for the thermal field and the thermal torque are
\be
\langle\Theta_{th}^{k}(t) \Theta_{th}^{j}(t')\rangle= 12 k_{B}T\eta V \delta_{kj}\delta(t-t'),
\ee 
\be
\langle H_{th}^{k}(t)H_{th}^{j}(t')\rangle=\frac{2k_{B}T \alpha}{\gamma \mu \mu_{0}^{2}}  \delta_{kj}\delta(t-t'),
\ee
where $k$ and $j$ index Cartesian vector components and $\delta_{kj}$ is the Kronecker delta function. \cite{engelmann2019predicting,chalifour2021magnetic,brown1963thermal} The stochastic fields/torques have a mean of zero, for example $\langle H_{th}^{k} \rangle = 0$. 

The equations of motion Eqs.~\eqref{LL} and \eqref{rotate} are integrated in time using 4th order Runge-Kutta integration~\cite{butcher1963coefficients} coded in Fortran. The timestep for rotation and Landau-Lifshitz motion of the moments is {0.5~ps.}

Two specific types of simulations were performed. The first investigates the Brownian relaxation only. This is achieved by locking the magnetic moment $\hat{m}$ to the easy axis $\hat{n}$, which forces the whole particle to rotate in the presence of the magnetic field. 
{After each timestep, $\hat{m}$ is aligned with $\hat{n}$ and the Landau-Lifshitz motion is ignored.}

The second type of simulation lets the magnetic moment move off the easy axis, allowing both Brownian and N\'eel relaxation to occur. 

Over the course of a simulation the total projection of the magnetization of all particles in the $z$ direction is recorded then averaged. This data is then fit to the decay function 
\be
\langle m_{z}(t) \rangle = (m_{0}-y_{0}) \exp(-t/\tau) + y_{0}
\label{fit}
\ee
where $\langle m_{z}(t) \rangle$ is the total average magnetization in the $z$ direction, and $m_{0}$, $y_{0}$, and $\tau$ are fitted parameters representing the initial averaged magnetization, the saturated value of the magnetic moments, and the characteristic relaxation time respectively. This approach is similar to our previous work~\cite{chalifour2021magnetic} to find N\'eel relaxation times.

\section{Results and Discussion}
\label{res}

To validate the numerical simulations, we first calculate the Brownian relaxation time in zero field, since there is a well-known expression for this time (Eq.~\eqref{ogbrown}).
In this case, the initial condition is that the magnetic moments are all aligned along the positive $z$ direction and then are allowed to thermally diffuse. (This is in contrast to simulations in a field which are started at random.) The net magnetic moment projected onto the $z$ axis is plotted versus time and Eq.~\eqref{fit} is fitted to the plot to find the Brownian relaxation time in zero applied field. In order to limit the stochastic influence in this simulation we increase the number of particles to {1000 in the 8~nm case and 500 particles for all other simulations.} Table~\ref{B=0table} compares the calculated relaxation times from our simulation to the analytic predictions found using Eq.~\eqref{ogbrown}, for a number of different particle radii.

\begin{table}[h!]
\begin{tabular}{|c|c|c|c|} 
 \hline
 Radius (nm) & Simulation ($\mu$s) & Eq.~\eqref{ogbrown} ($\mu$s) & \% Error\\ [0.5ex] 
 \hline\hline
 8 & 5.05  & 5.4 & 6.51 \\ 
 \hline
 10 & 10.56 & 10.6 & 0.38 \\
 \hline
 12 & 18.48 & 18.4 & 0.43 \\
 \hline
 14 & 28.75 & 29.1 & 1.21 \\
 \hline
  15 & 37.01 & 35.9 & 3.09 \\
 \hline
 16 & 43.15 & 43.5 & 0.80\\ [0.5ex] 
 \hline
\end{tabular}
\caption{Calculated Brownian relaxation time from our simulations with zero applied field, $B$=0~mT, compared to the analytical expression for the Brownian relaxation time (Eq.~\eqref{ogbrown}). Results are shown for {1000 particles for 8~nm and 500 for all other} radii and $T=300$~K. Other particle parameters are quoted in the first paragraph of Sec.~\ref{Sim}.}
\label{B=0table}
\end{table}

One can see in Table~\ref{B=0table} that the relaxation time found via particle-level simulation matches well with the predictions of analytic Eq.~\eqref{ogbrown}. The difference is 3\% or less for all values except for the smallest particles (8~nm radius) where the ratio of the simulation timestep to the relaxation time is larger. This gives us confidence in the numerical simulation results.

Note that the simulation results are {limited by the number of particles that are used (1000 or 500) and by the timestep used (0.5~ps).} However, we have checked that re-running simulations {for more particles or smaller timesteps} does not change the result markedly. Also, the relaxation time has some uncertainty associated with fitting the parameters of Eq.~\eqref{fit} which is displayed in the simulations column of Table~\ref{B=0table}.

\begin{figure}
\centering
\includegraphics[width = 0.9\columnwidth]{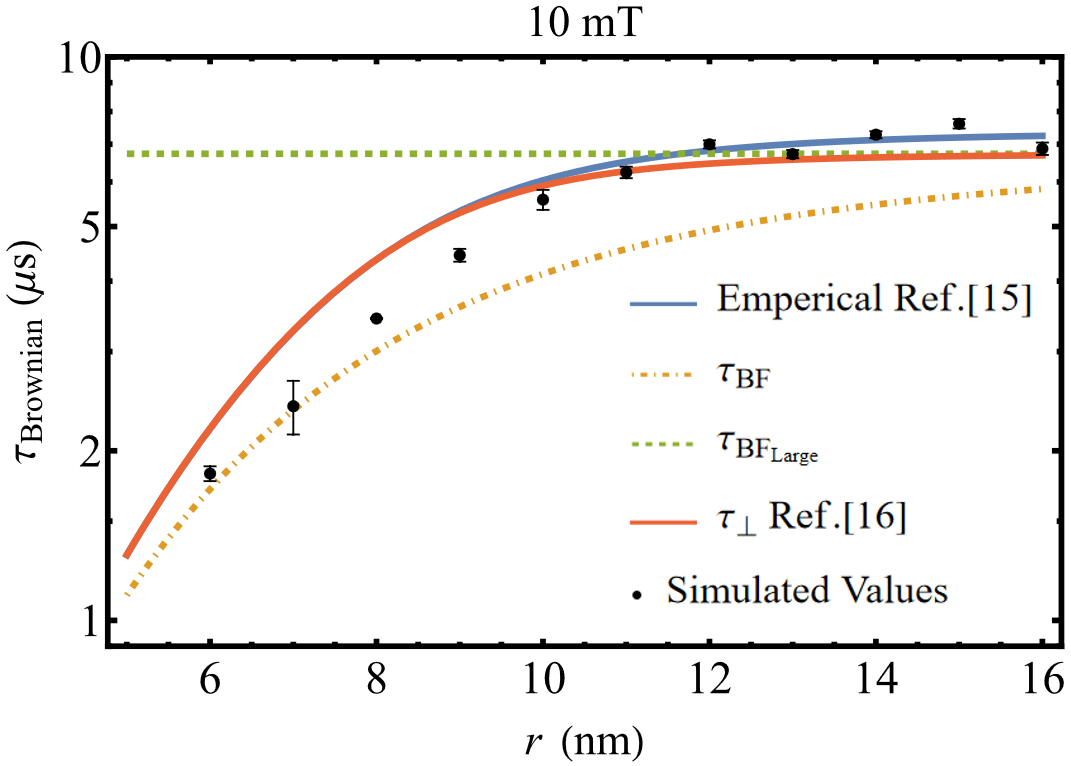}
\caption{Simulated Brownian relaxation time as a function of radius (black dots), in the case of a 10~mT applied field and $T=300$~K. This is compared to results found using Eq.~\eqref{Brownianfield} (orange dot-dash) and Eq.~\eqref{BrownianfieldApprox} (green dashed), along with an empirical expression found in literature \cite{yoshida2009simulation} (solid line) {and $\tau_{\perp}$ (red solid line) from} \cite{martsenyuk1974kinetics}. Particle parameters are quoted at the start of Sec.~\ref{Sim}.}
\label{Brownian}
\end{figure}

With the accuracy of the simulations checked in zero field, we now look at the Brownian relaxation time in a magnetic field.
Fig.~\ref{Brownian} presents the Brownian relaxation time in the presence of a 10~mT field, as a function of particle radius $r$.
Note that in the case of small radii ($r<7$~nm), 1000 particles -- rather than the usual 500 -- are simulated in order to obtain a smoother average moment versus time. This is because thermal fluctuations are a bigger contribution to the total torque on small particles, meaning the moments have a more random trajectory than for larger particles. Results of various calculations are shown, including: our results from simulations (dots), our analytic expression (Eq.~\eqref{Brownianfield}, dot-dashed), our analytic large-radii approximation (Eq.~\eqref{BrownianfieldApprox}, dashed){,} an empirically found expression from the literature (Ref.~\cite{yoshida2009simulation}, {blue} solid line){, and $\tau_{\perp}$ (red solid line) from} \cite{martsenyuk1974kinetics}. Again, the relaxation time from simulated data was found by fitting the total magnetization of the MNPs as a function of time to Eq.\eqref{fit}. {Error bars for the simulated values were found by running the simulation twice to obtain variations in the stochastic forces and then the standard error was found for the fitted characteristic time.} Note that there is no N\'eel relaxation at this stage but this will be considered later in the section.

As a first point, notice that the relaxation times in this moderate magnetic field of 10~mT (100~Oe) are much shorter than in zero field. For example, consider the 16~nm particle radius. In zero field (Table~\ref{B=0table}) the Brownian relaxation time is 43.9~$\mu$s, whereas in 10~mT (see Fig.~\ref{Brownian}) it is around 7~$\mu$s. This is a change of an order of magnitude, due to the magnetic field driving the relaxation, which is therefore faster than diffusive relaxation. This large change shows that using expressions that are developed for zero field are inappropriate in applied field strengths that are typical for many experiments and MNP applications.

Fig.~\ref{Brownian} reveals that our approximate relaxation time (Eq.~\eqref{Brownianfield}, dot-dashed) displays the same qualitative trend as the simulated data (dots), but slightly underestimates the relaxation time, {apart from at small radii $r<8$~nm}.
This supports our assertion in Sec.~\ref{meth} that ignoring the nonlinear term in the Fokker-Planck equation leads to an approximate Brownian relaxation time which is relatively accurate.
Moreover, {the situation where our approximation works best (small radii) is where the magnetic moments reorient the least, meaning most are near the equator and the nonlinear term that we neglected is small.}

Our approximate, large-radii form of the field-dependent Brownian relaxation time (Eq.~\eqref{BrownianfieldApprox}, flat dashed line) matches well with the simulated data for nanoparticle radii above 12~nm. This is as expected from our discussion in Sec.~\ref{meth}, since the Zeeman energy dominates over thermal energy in this limit (ie. $2k_{B} T << \mu B$). {As mentioned earlier in Sec. II, our large radii limit coincides with the large radii limit given in Ref.}~\cite{martsenyuk1974kinetics} {(red, solid line).} 

{For intermediate radii ($8<r<12$~nm)}, the empirically derived expression from Ref.~\cite{yoshida2009simulation} {(blue, solid line) and $\tau_{\perp}$ from} \cite{martsenyuk1974kinetics} {(red, solid line)} provide a more accurate fit to the simulated data (dots). This is most likely due to the fact that our analytic expressions are found by removing the nonlinear contribution of the differential equation. {Note the similarities between the empirical expression from Ref.}~\cite{yoshida2009simulation} {and $\tau_{\perp}$ from} \cite{martsenyuk1974kinetics}. {The two expression only start to visually deviate from each other above a radius of $10$~nm.}

We can also compare results of our analytic expression Eq.~\eqref{Brownianfield} to numerical calculations done by Deissler \emph{et al.} \cite{deissler2014dependence}, which utilizes eigenvalue calculations to find a field dependant Brownian relaxation time. To compare directly we use material parameters to match with Ref.~\cite{deissler2014dependence}, namely $r=10$~nm, $T=300$~K, $\eta=1.0049$~mPa~s, $K=20$~kJ/m$^{3}$, $M_{sat}=474$~kA/m, $\alpha=0.1$, and $\gamma=1.75\times10^{11}$~rad/s~T at 20~mT applied field. Where Deissler \emph{et al.} calculates the Brownian relaxation time to be $0.37~\mu$s, Eq.~\eqref{Brownianfield} yields $0.53~\mu$s. This is  a similar value utilizing a simpler, analytic expression.

It is important to note that as the radii decreases, the N\'eel relaxation mechanism generally dominates the dynamics of the magnetic moment \cite{deissler2013brownian,brown1962magnetostatic}, so in reality one would have to include both N\'eel and Brownian mechanisms to obtain the correct \emph{total} characteristic relaxation time. However, Fig.~\ref{Brownian} is presented to show the validity of various analytic expressions for the Brownian relaxation time \emph{in a field}, before considering N\'eel dynamics.

\begin{figure}
\centering
\includegraphics[width = 0.90\columnwidth]{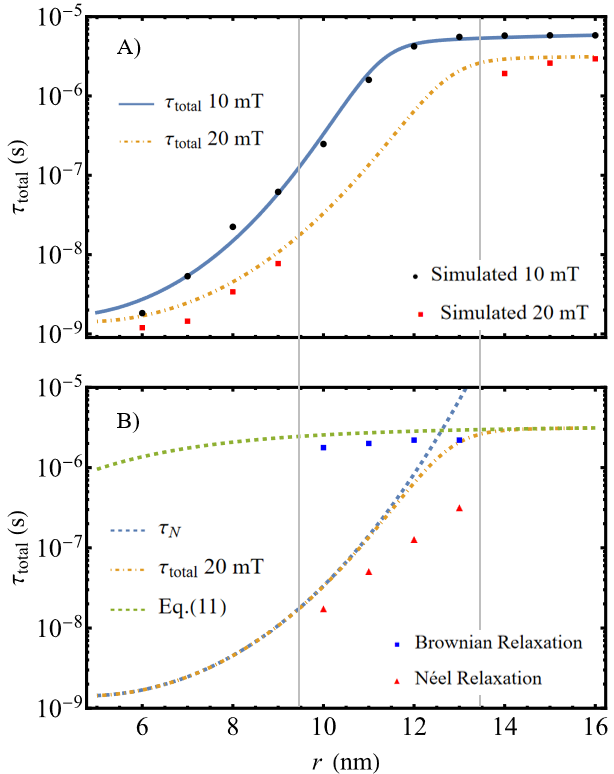}
\caption{ Calculated total relaxation time in an applied field, including both Brownian and N\'eel relaxation mechanisms, as a function of particle radius. {In panel (a) t}he results of simulation are shown for 10~mT (black dots) and 20~mT (red squares). The lines are predictions for the  total relaxation time $\tau_{\textrm{total}}$ found by reciprocally adding N\'eel and Brownian relaxation times, as described in the main text Eq.~\eqref{totaltime}. The $10$~mT result (blue solid line) and $20$~mT result (orange dot-dashed line) are shown. Note the region between 9--14~nm {(denoted by the gray vertical lines)} without simulated results in the case of 20~mT, due to the inability to  fit with a single exponential function. {Panel (b) shows the simulated results when fitting with two exponential functions. The slower characteristic time (blue squares) is close to the Brownian relaxation time in a field predicted by Eq.}~\eqref{Brownianfield} {(green dashed line) and the faster characteristic time (red triangles) is attributed to the N\'eel relaxation time $\tau_{N}$ (blue dashed line).}}
\label{Total}
\end{figure}

In Fig.~\ref{Total} we include the N\'eel dynamics along with Brownian dynamics in our simulations and plot the total relaxation time versus particle radius. The results of fitting to simulations are given (dots), along with an analytic approximation (lines). Two field values are considered: 10~mT (black dots and solid line), plus 20~mT (red squares and dot-dashed line). The analytic approximation needs some explanation. The total relaxation time can be approximated by reciprocally adding the {field-dependant Brownian relaxation time} we found and verified {(Eq.}~\eqref{Brownianfield}) and an expression for the field-dependent N\'eel relaxation time found in our previous work (equation number 23 of Ref.~\cite{chalifour2021magnetic}).  As discussed when developing Eq.~\eqref{timecombine}, this approach is motivated by the fact that the shortest timescale dominates the overall relaxation time \cite{rosensweig2002heating}. The total characteristic relaxation time $\tau_{\textrm{total}}$ is given by
\be
\frac{1}{\tau_{\textrm{total}}} = \frac{1}{\tau_{N}} + \frac{1}{\tau_{BF~\textrm{}}},
\label{totaltime}
\ee
where $\tau_{N}$ is the N\'eel relaxation time for a system with random alignment of easy axes, in an applied field.

In Fig.~\ref{Total}, the 10~mT simulated results (dots) match extremely well with our analytic result (solid line). This indicates that the use of Eq.~\eqref{totaltime} to predict the relaxation time of single-domain, noninteracting magnetic nanoparticles \emph{in a magnetic field} is robust. Note, however, that the times are on a logarithmic scale in Fig.~\ref{Total}, compared to Fig.~\ref{Brownian} where the scale was linear. 

In the case of the 20~mT field, only the simulated results for small particles (between 6--9~nm radius) and larger particles (14--16~nm radius) are {shown} on Fig.~\ref{Total}{(a)}. The region between 9--14 nm is not shown because the simulation results for $\langle m_z \rangle$ as a function of time could not be fitted with one single exponential function (Eq.~\eqref{fit}). This is because of the phenomenon mentioned in the introduction, where N\'eel relaxation happens first then Brownian relaxation occurs later \cite{ota2016rotation}, rather than both happening simultaneously as Eq.~\eqref{totaltime} assumes. 

\begin{figure}
\centering
\includegraphics[width = 0.9\columnwidth]{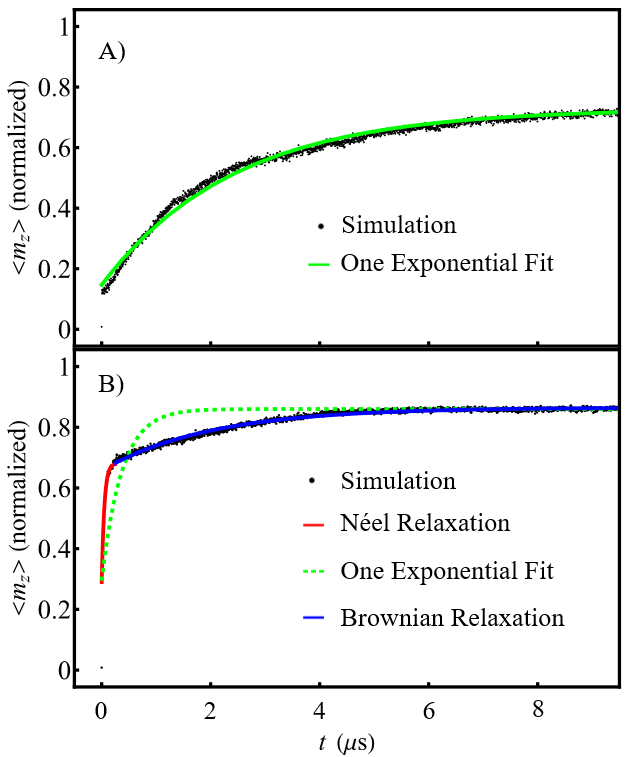}
\caption{The average projection of magnetization onto the $z$ axis as a function of time produced by our particle-level simulation (black dots) for a system of 500 11-nm-radius particles, {in (a) 10~mT and (b) 20~mT applied field and at $T=300$~K. If one uses a single exponential in the form of} Eq.~\eqref{fit} {for the 10~mT applied field case (green solid line) it does accurately describe how the magnetization changes over time. However, in the case of 20~mT applied field (green dashed line) it does not accurately describe the magnetization dynamics.} One must break up the data into two regions associated with N\'eel ($t<0.2~\mu$s, red solid line) and Brownian (longer times, blue solid line) relaxation in order to find two separate values of the fitting parameter $\tau$.} 
\label{fitting}
\end{figure}

To demonstrate this separation of N\'eel and Brownian timescales, in Fig.~\ref{fitting} the total, normalized magnetization versus time is plotted for one such scenario. Fig.~\ref{fitting} shows the average magnetization in the $z$ direction of all 500 particles in a simulation, with radius of 11~nm and in a {10~mT applied field (a) and 20~mT applied field (b) at $T=300$~K.} {In the case of the 10~mT applied field (a) the fitted exponential function (green solid line) does accurately describe how the magnetization changes over time and a single characteristic time can be found. When the applied field is increased to 20~mT (b), one can see fitting the data with a single exponential in the form of} Eq.~\eqref{fit} (green dashed line) does not accurately represent the dynamics of the average magnetisation. However, if one breaks up the data into two regions, one can find separate N\'eel ($t<0.2~\mu$s, red solid line) and Brownian (longer times, blue solid line) relaxation times with vastly different fitting parameters $\tau$. {In Fig.}~\ref{Total}{(b) we follow this procedure for all simulations of particle radius 10--13~nm in 20~mT applied field and find the short time fit (red triangles) and the long time fit (blue squares). This data is then compared to the predicted total relaxation time Eq.}~\eqref{totaltime} {in a 20~mT applied field (orange dot-dashed line) alongside the Brownian relaxation time in a field Eq.}~\eqref{Brownianfield} {(green dashed line) and $\tau_{N}$ (blue dashed line). Here one can see the slower fitted times correlate to the Brownian relaxation time in a field and the faster fitted times to the N\'eel relaxation time.}  

From Fig.~\ref{fitting}{(b)}, the N\'eel relaxation time (short time fit) is 0.052~$\mu$s, the Brownian relaxation time (long time fit) is 2.0~$\mu$s, and the \emph{incorrect} single exponential fit yields 0.18~$\mu$s. Using Eq.~\eqref{totaltime} to predict the total relaxation time gives 0.14~$\mu$s, but this number is unphysical since the short and long timescale dynamics are decoupled. In this case, therefore, it seems ill-advised to use Eq.~\eqref{totaltime} to make predictions for the total relaxation time.

It is intriguing that this separation of short- and long-time relaxation (the emergence of two relaxation times) only occurs for certain sets of parameters, such as between 9--14~nm radius and at 20~mT applied field for the magnetite particles considered in Fig.~\ref{Total}. We explored why this occurs for 20~mT but not for 10~mT in the system investigated here. Although we could not come up with a way to predict the occurrence of the two-timescale behavior, we did find a way to characterize the dynamics and associate it with magnetic energy barriers.

\begin{figure}
\centering
\includegraphics[width = 0.9\columnwidth]{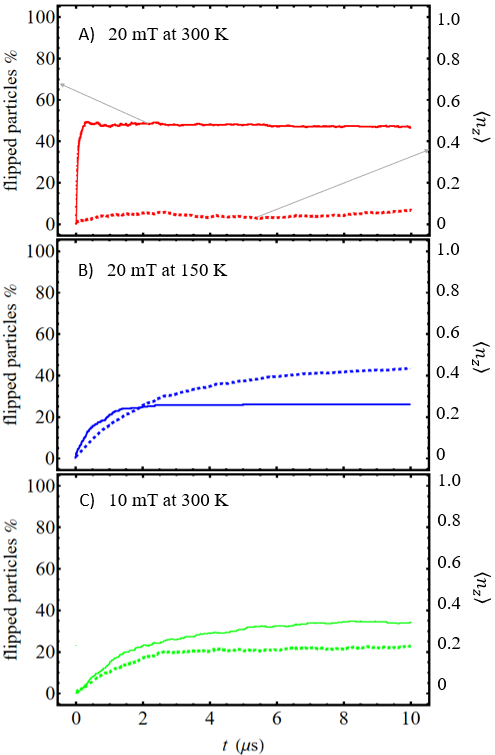}
\caption{The percentage of particles that have undergone a ``flip" over an energy barrier (solid lines) plus the average projection of particles' easy axes $\langle n_z \rangle$ (dashed line) as a function of time, for three simulations. Panel (a) is for 20~mT at 300~K, panel (b) is the same field but at 150~K, and panel (c) is 10~mT at 300~K. These three examples are chosen to show that either reducing the temperature or the applied field strength leads to both N\'eel (flips) and Brownian (easy axis rotation) occurring on the same timescales.
}
\label{flipping and easy}
\end{figure}

When two relaxation timescales are observed, it is because at very short timescales there are many excursions of the magnetic moment over an internal energy barrier. In contrast, when there is a single relaxation time, this ``flipping" of the magnetic moment over a barrier happens on timescales that are comparable with the timescale for the physical Brownian rotation of the particles. This is shown in Fig.~\ref{flipping and easy} which shows three different simulation dynamics for the magnetic moments of 500 11-nm-radius particles in (a) 20~mT field at 300~K, (b) 20~mT field at 150~K, and (c) 10~mT field at 300~K. The solid lines show the percent of particles that have experienced a magnetic ``flip" over an internal energy barrier and the dashed line shows the mean projection of a particle's easy axis in the $z$ direction $\langle n_z \rangle$ (physical rotation), as a function of time. {Note that initially $\langle n_z \rangle=0$ in all panels because the particles' easy axes are randomly oriented. Since the easy axis can be defined in two different ways (with a positive or negative component in $z$), then our choice to add up all the easy axes in this way is just one option. It is chosen so that a lack of orientational order corresponds to $\langle n_z \rangle=0$. With our definition, any $\langle n_z \rangle>0$ shows a preference for easy axes to align with the applied field direction. Here, the magnitude of $\langle n_z \rangle$ is less important than the timescale of its approach to a steady-state value.}

As one can see {in Fig.}~\ref{flipping and easy} for (b) 20~mT at 150~K and (c) 10~mT at 300~K, both {internal magnetization flipping and easy axis rotation} dynamics occur on similar timescales, {on the order of a few microseconds. In contrast} for (a) 20~mT at 300~K the flipping occurs in under 0.2~$\mu$s while the Brownian rotations occur over tens of microseconds. In fact, roughly 50~\% of particles have flipped in under 0.2~$\mu$s, which is the maximum expected number since in the random starting configuration, half of the particles are expected to have moments with a $z$ component opposite the applied field direction. Fig.~\ref{flipping and easy} indicates that the strength of the applied field and the temperature play large roles in determining the emergence of the two separate  relaxation times. 

The flipping of moments over a magnetic energy barrier is dependent on the size of the energy barrier compared to thermal energy fluctuations. For an applied field of 20~mT, the energy barriers are reduced in size compared to 10~mT, so the increased rate of flipping seen in Fig.~\ref{flipping and easy}(a) can be rationalized this way. We have also confirmed through simulations that if the system of magnetic nanoparticles is cooled to, say, 150~K, then the energy barrier cannot be traversed easily and the dynamics collapses to having a single relaxation time (compare panels (a) and (b)). Although the emergence of two relaxation timescales -- rather than one -- seems to be related to energy barriers, we were unable to accurately predict using analytic tools when this behavior is expected to occur. In all cases, the energy barrier for reversal (estimated using Stoner-Wohlfarth theory \cite{tannous2008stoner}) is on the order of a few times $k_B T$.

Those researchers using magnetic nanoparticle dynamics induced by applied fields for applications in, for example, magnetic hyperthermia or magnetic particle imaging, should be aware of the two-timescale behavior. For example, an incorrect estimation of the relaxation time could result in the estimation of heat produced by a certain MNP in a certain magnetic field to be drastically off. To accurately represent the magnetization dynamics and therefore the work done by a magnetic field and possibly the heat produced, one would need to estimate the short and long timescales separately.

\section{conclusion}

An approximate analytic expression for the field-dependent Brownian magnetic relaxation time $\tau_{BF }$ of a system of magnetic nanoparticles is derived, namely Eq.~\eqref{Brownianfield}. In addition, Eq.~\eqref{Brownianfield} can be further approximated in the large-particle limit where $2 k_B T << M_s V B$, to give Eq.~\eqref{BrownianfieldApprox} for the approximate relaxation time $\tau_{BF large}$. These expressions are validated through particle-level dynamics simulations done on a collection of uniformly sized, single domain, non-interacting magnetic nanoparticles. A field-dependent expression that appears in the literature~\cite{yoshida2009simulation} matches well with simulations over a broader range of parameters, but is an empirical fit, rather than being derived from thermal relaxation equations.
The expression for the field-dependent Brownian relaxation time is combined with an expression for the field-dependent N\'eel relaxation time given by Ref.~\cite{chalifour2021magnetic} in Eq.~\eqref{totaltime} to provide a field-dependent total magnetic relaxation time. Again, its validity is tested against particle-level simulations. For a 10~mT applied field, the analytic expression matches well with the results of simulations across a broad range of magnetite nanoparticle sizes. Since the Brownian relaxation dominates at large particle sizes, the simple approximation given by Eq.~\eqref{BrownianfieldApprox} works well in helping to determine the total relaxation time via Eq.~\eqref{totaltime}.

However, for magnetite particles in a 20~mT field, simulations showed that a single relaxation time was not appropriate in describing the net magnetization dynamics of an ensemble of particles. At 300~K and for particles with radii between 9 and 14~nm, this was the case. Therefore, one must be cautious in using Eq.~\eqref{totaltime} to estimate the field-dependent total magnetic relaxation time as it is not always valid. We found -- as did others \cite{ota2016rotation} -- that many particle moments flip over energy barriers in sub-microsecond timescales, before a much slower Brownian rotation takes place. This decoupling of the relaxation mechanisms is related to the size of energy barriers compared to thermal fluctuations, and so can be turned off by reducing the applied field or the temperature. However, a way to predict when it will occur is a topic for future work.

Here, dipolar interactions between particles have been ignored, although they were considered in the past.~\cite{dormann1997magnetic} Future work could adapt the analytic relaxation time expressions developed here for the case of weak interactions, since they can have important consequences on magnetic behavior.~\cite{chantrell2000calculations,nunes2005role} Eq.~\eqref{totaltime} should only be used for dilute systems without this extension. Additionally, interacting particles can form chains of particles which vastly affects the rotational dynamics of the particles in an applied field. \cite{kim2015relaxation, anderson2021simulating} 

Another assumption made here is that the particles are perfectly spherical and all have the same size. In reality, the particles are polydisperse, meaning the characteristic relaxation time would vary from particle to particle and the net relaxation time of the system should be estimated accordingly.~\cite{livesey2018beyond,respaud1998surface} {As mentioned previously, magnetite has a cubic anisotropy which is modeled as uniaxial in this work to compare to other calculations of the relaxation times.}~\cite{yoshida2009simulation,martsenyuk1974kinetics} {In the future we can extend the study of relaxation times to cubic anisotropy.}

Finally, we note that here the hydrodynamic volume was assumed to be the same as the magnetic volume of the particles. For magnetic particles with ligands on their surface, the hydrodynamic radius can be much larger than the magnetic radius, which would change the derivation of Eqs.~\eqref{Brownianfield} and \eqref{BrownianfieldApprox}.

\begin{acknowledgments}
 We thank Dr Artek R.~Chalifour for useful discussions. J.C.D. acknowledges support from a UCCS Mentored Doctoral Fellowship. J.C.D. and K.L.L. acknowledge funding from National Science Foundation Award DMR-1808412. The authors thank the UCCS Biofrontiers Center for support.
\end{acknowledgments}

\appendix
\section*{Appendix: Analytic Brownian relaxation time expressions from literature}

{Martsenyuk \emph{et al.} }\cite{martsenyuk1974kinetics}{ derived an analytic expression for the Brownian relaxation time of the ensemble magnetization components both transverse and longitudinal to an applied field. It is the transverse component which corresponds to our result for $\tau_{BF}$, since the moments are predominantly located near the equator before partially aligning in the transverse direction of the field, along the $+z$ pole. Their expression is }
\be
\tau_{\perp}= \frac{2L(\xi)}{\xi-L(\xi)} \tau_B
\label{marty}
\ee
where
\be
\xi= \frac{\mu B}{k_{B}T},~~~~~~\mu= M_{s}V,
\label{xi}
\ee
and $L(\xi)$ is the Langevin function, namely
\be
L(\xi) = \coth(\xi) - \frac{1}{\xi}.
\ee
Here, the Brownian relaxation time in \emph{zero field} $\tau_B$ is given in Eq.~\eqref{ogbrown}. The pre-factor in Eq.~\eqref{marty} therefore corresponds to a unitless correction that depends on the applied field strength. The result of Eq.~\eqref{marty} is given by the solid red line in Fig.~\ref{Brownian}.

{For large particles and/or large field strength (i.e. $\xi >> 1$), Eq.}~\eqref{marty}{ reduces to}
\be
\tau_{\perp}= \frac{2\tau_{B}}{\xi}.
\label{martyLarge}
\ee
This result precisely matches our Eq.~\eqref{BrownianfieldApprox} for $\tau_{BF \textrm{large}}$, which is derived in the same large particle limit.

{Another expression exists for the Brownian relaxation time in a field, provided by Yoshida and Enpuku}~\cite{yoshida2009simulation}{. This was found as an empirical fit to computer simulation results. The empirical formula is}
\be
\tau_{\textrm{eff}}= \frac{\tau_{B}}{\sqrt{1+0.21\xi^{2}}},
\ee
where $\xi$ is defined as in Eq.~\eqref{xi}, and $\tau_B$ is given in Eq.~\eqref{ogbrown}. This result is shown by the solid blue line in Fig.~\ref{Brownian}.

\bibliography{bib}

\begin{thebibliography}{46}%
\makeatletter
\providecommand \@ifxundefined [1]{%
 \@ifx{#1\undefined}
}%
\providecommand \@ifnum [1]{%
 \ifnum #1\expandafter \@firstoftwo
 \else \expandafter \@secondoftwo
 \fi
}%
\providecommand \@ifx [1]{%
 \ifx #1\expandafter \@firstoftwo
 \else \expandafter \@secondoftwo
 \fi
}%
\providecommand \natexlab [1]{#1}%
\providecommand \enquote  [1]{``#1''}%
\providecommand \bibnamefont  [1]{#1}%
\providecommand \bibfnamefont [1]{#1}%
\providecommand \citenamefont [1]{#1}%
\providecommand \href@noop [0]{\@secondoftwo}%
\providecommand \href [0]{\begingroup \@sanitize@url \@href}%
\providecommand \@href[1]{\@@startlink{#1}\@@href}%
\providecommand \@@href[1]{\endgroup#1\@@endlink}%
\providecommand \@sanitize@url [0]{\catcode `\\12\catcode `\$12\catcode `\&12\catcode `\#12\catcode `\^12\catcode `\_12\catcode `\%12\relax}%
\providecommand \@@startlink[1]{}%
\providecommand \@@endlink[0]{}%
\providecommand \url  [0]{\begingroup\@sanitize@url \@url }%
\providecommand \@url [1]{\endgroup\@href {#1}{\urlprefix }}%
\providecommand \urlprefix  [0]{URL }%
\providecommand \Eprint [0]{\href }%
\providecommand \doibase [0]{https://doi.org/}%
\providecommand \selectlanguage [0]{\@gobble}%
\providecommand \bibinfo  [0]{\@secondoftwo}%
\providecommand \bibfield  [0]{\@secondoftwo}%
\providecommand \translation [1]{[#1]}%
\providecommand \BibitemOpen [0]{}%
\providecommand \bibitemStop [0]{}%
\providecommand \bibitemNoStop [0]{.\EOS\space}%
\providecommand \EOS [0]{\spacefactor3000\relax}%
\providecommand \BibitemShut  [1]{\csname bibitem#1\endcsname}%
\let\auto@bib@innerbib\@empty
\bibitem [{\citenamefont {Shasha}\ and\ \citenamefont {Krishnan}(2021)}]{shasha2021nonequilibrium}%
  \BibitemOpen
  \bibfield  {author} {\bibinfo {author} {\bibfnamefont {C.}~\bibnamefont {Shasha}}\ and\ \bibinfo {author} {\bibfnamefont {K.~M.}\ \bibnamefont {Krishnan}},\ }\bibfield  {title} {\bibinfo {title} {{N}onequilibrium {D}ynamics of {M}agnetic {N}anoparticles with {A}pplications in {B}iomedicine},\ }\href@noop {} {\bibfield  {journal} {\bibinfo  {journal} {Adv. Mater.}\ }\textbf {\bibinfo {volume} {33}},\ \bibinfo {pages} {1904131} (\bibinfo {year} {2021})}\BibitemShut {NoStop}%
\bibitem [{\citenamefont {Tang}\ and\ \citenamefont {Lo}(2013)}]{tang2013magnetic}%
  \BibitemOpen
  \bibfield  {author} {\bibinfo {author} {\bibfnamefont {S.~C.}\ \bibnamefont {Tang}}\ and\ \bibinfo {author} {\bibfnamefont {I.~M.}\ \bibnamefont {Lo}},\ }\bibfield  {title} {\bibinfo {title} {{M}agnetic nanoparticles: {E}ssential factors for sustainable environmental applications},\ }\href@noop {} {\bibfield  {journal} {\bibinfo  {journal} {Water Res.}\ }\textbf {\bibinfo {volume} {47}},\ \bibinfo {pages} {2613} (\bibinfo {year} {2013})}\BibitemShut {NoStop}%
\bibitem [{\citenamefont {Deatsch}\ and\ \citenamefont {Evans}(2014)}]{deatsch2014heating}%
  \BibitemOpen
  \bibfield  {author} {\bibinfo {author} {\bibfnamefont {A.~E.}\ \bibnamefont {Deatsch}}\ and\ \bibinfo {author} {\bibfnamefont {B.~A.}\ \bibnamefont {Evans}},\ }\bibfield  {title} {\bibinfo {title} {{H}eating efficiency in magnetic nanoparticle hyperthermia},\ }\href@noop {} {\bibfield  {journal} {\bibinfo  {journal} {J. Magn. Magn. Mater.}\ }\textbf {\bibinfo {volume} {354}},\ \bibinfo {pages} {163} (\bibinfo {year} {2014})}\BibitemShut {NoStop}%
\bibitem [{\citenamefont {Torres}\ \emph {et~al.}(2019)\citenamefont {Torres}, \citenamefont {Lima~Jr}, \citenamefont {Calatayud}, \citenamefont {Sanz}, \citenamefont {Ibarra}, \citenamefont {Fern{\'a}ndez-Pacheco}, \citenamefont {Mayoral}, \citenamefont {Marquina}, \citenamefont {Ibarra},\ and\ \citenamefont {Goya}}]{torres2019relevance}%
  \BibitemOpen
  \bibfield  {author} {\bibinfo {author} {\bibfnamefont {T.~E.}\ \bibnamefont {Torres}}, \bibinfo {author} {\bibfnamefont {E.}~\bibnamefont {Lima~Jr}}, \bibinfo {author} {\bibfnamefont {M.~P.}\ \bibnamefont {Calatayud}}, \bibinfo {author} {\bibfnamefont {B.}~\bibnamefont {Sanz}}, \bibinfo {author} {\bibfnamefont {A.}~\bibnamefont {Ibarra}}, \bibinfo {author} {\bibfnamefont {R.}~\bibnamefont {Fern{\'a}ndez-Pacheco}}, \bibinfo {author} {\bibfnamefont {A.}~\bibnamefont {Mayoral}}, \bibinfo {author} {\bibfnamefont {C.}~\bibnamefont {Marquina}}, \bibinfo {author} {\bibfnamefont {M.~R.}\ \bibnamefont {Ibarra}},\ and\ \bibinfo {author} {\bibfnamefont {G.~F.}\ \bibnamefont {Goya}},\ }\bibfield  {title} {\bibinfo {title} {{T}he relevance of {B}rownian relaxation as power absorption mechanism in {M}agnetic {H}yperthermia},\ }\href@noop {} {\bibfield  {journal} {\bibinfo  {journal} {Sci. Rep.}\ }\textbf {\bibinfo {volume} {9}},\ \bibinfo {pages} {3992} (\bibinfo {year} {2019})}\BibitemShut {NoStop}%
\bibitem [{\citenamefont {Rajan}\ \emph {et~al.}(2020)\citenamefont {Rajan}, \citenamefont {Sharma},\ and\ \citenamefont {Sahu}}]{rajan2020assessing}%
  \BibitemOpen
  \bibfield  {author} {\bibinfo {author} {\bibfnamefont {A.}~\bibnamefont {Rajan}}, \bibinfo {author} {\bibfnamefont {M.}~\bibnamefont {Sharma}},\ and\ \bibinfo {author} {\bibfnamefont {N.~K.}\ \bibnamefont {Sahu}},\ }\bibfield  {title} {\bibinfo {title} {{A}ssessing magnetic and inductive thermal properties of various surfactants functionalised {F}e$_{3}${O}$_{4}$ nanoparticles for hyperthermia},\ }\href@noop {} {\bibfield  {journal} {\bibinfo  {journal} {Sci. Rep.}\ }\textbf {\bibinfo {volume} {10}},\ \bibinfo {pages} {15045} (\bibinfo {year} {2020})}\BibitemShut {NoStop}%
\bibitem [{\citenamefont {Giri}\ \emph {et~al.}(2005)\citenamefont {Giri}, \citenamefont {Pradhan}, \citenamefont {Sriharsha},\ and\ \citenamefont {Bahadur}}]{giri2005preparation}%
  \BibitemOpen
  \bibfield  {author} {\bibinfo {author} {\bibfnamefont {J.}~\bibnamefont {Giri}}, \bibinfo {author} {\bibfnamefont {P.}~\bibnamefont {Pradhan}}, \bibinfo {author} {\bibfnamefont {T.}~\bibnamefont {Sriharsha}},\ and\ \bibinfo {author} {\bibfnamefont {D.}~\bibnamefont {Bahadur}},\ }\bibfield  {title} {\bibinfo {title} {{P}reparation and investigation of potentiality of different soft ferrites for hyperthermia applications},\ }\href@noop {} {\bibfield  {journal} {\bibinfo  {journal} {J. Appl. Phys.}\ }\textbf {\bibinfo {volume} {97}},\ \bibinfo {pages} {10Q916} (\bibinfo {year} {2005})}\BibitemShut {NoStop}%
\bibitem [{\citenamefont {Croft}\ \emph {et~al.}(2012)\citenamefont {Croft}, \citenamefont {Goodwill},\ and\ \citenamefont {Conolly}}]{croft2012relaxation}%
  \BibitemOpen
  \bibfield  {author} {\bibinfo {author} {\bibfnamefont {L.~R.}\ \bibnamefont {Croft}}, \bibinfo {author} {\bibfnamefont {P.~W.}\ \bibnamefont {Goodwill}},\ and\ \bibinfo {author} {\bibfnamefont {S.~M.}\ \bibnamefont {Conolly}},\ }\bibfield  {title} {\bibinfo {title} {{R}elaxation in {X}-{S}pace {M}agnetic {P}article {I}maging},\ }\href@noop {} {\bibfield  {journal} {\bibinfo  {journal} {IEEE T. Med. Imaging}\ }\textbf {\bibinfo {volume} {31}},\ \bibinfo {pages} {2335} (\bibinfo {year} {2012})}\BibitemShut {NoStop}%
\bibitem [{\citenamefont {Deissler}\ \emph {et~al.}(2013)\citenamefont {Deissler}, \citenamefont {Martens}, \citenamefont {Wu},\ and\ \citenamefont {Brown}}]{deissler2013brownian}%
  \BibitemOpen
  \bibfield  {author} {\bibinfo {author} {\bibfnamefont {R.~J.}\ \bibnamefont {Deissler}}, \bibinfo {author} {\bibfnamefont {M.~A.}\ \bibnamefont {Martens}}, \bibinfo {author} {\bibfnamefont {Y.}~\bibnamefont {Wu}},\ and\ \bibinfo {author} {\bibfnamefont {R.}~\bibnamefont {Brown}},\ }\bibfield  {title} {\bibinfo {title} {{B}rownian and {N}{\'e}el relaxation times in magnetic particle dynamics},\ }in\ \href@noop {} {\emph {\bibinfo {booktitle} {2013 International Workshop on Magnetic Particle Imaging (IWMPI)}}}\ (\bibinfo {organization} {IEEE},\ \bibinfo {year} {2013})\ pp.\ \bibinfo {pages} {1--1}\BibitemShut {NoStop}%
\bibitem [{\citenamefont {Rosensweig}(2002)}]{rosensweig2002heating}%
  \BibitemOpen
  \bibfield  {author} {\bibinfo {author} {\bibfnamefont {R.~E.}\ \bibnamefont {Rosensweig}},\ }\bibfield  {title} {\bibinfo {title} {{H}eating magnetic fluid with alternating magnetic field},\ }\href@noop {} {\bibfield  {journal} {\bibinfo  {journal} {J. Magn. Magn. Mater.}\ }\textbf {\bibinfo {volume} {252}},\ \bibinfo {pages} {370} (\bibinfo {year} {2002})}\BibitemShut {NoStop}%
\bibitem [{\citenamefont {Wu}\ \emph {et~al.}(2019)\citenamefont {Wu}, \citenamefont {Zhang}, \citenamefont {Steinberg}, \citenamefont {Qu}, \citenamefont {Huang}, \citenamefont {Cheng}, \citenamefont {Bliss}, \citenamefont {Du}, \citenamefont {Rao}, \citenamefont {Song} \emph {et~al.}}]{wu2019review}%
  \BibitemOpen
  \bibfield  {author} {\bibinfo {author} {\bibfnamefont {L.~C.}\ \bibnamefont {Wu}}, \bibinfo {author} {\bibfnamefont {Y.}~\bibnamefont {Zhang}}, \bibinfo {author} {\bibfnamefont {G.}~\bibnamefont {Steinberg}}, \bibinfo {author} {\bibfnamefont {H.}~\bibnamefont {Qu}}, \bibinfo {author} {\bibfnamefont {S.}~\bibnamefont {Huang}}, \bibinfo {author} {\bibfnamefont {M.}~\bibnamefont {Cheng}}, \bibinfo {author} {\bibfnamefont {T.}~\bibnamefont {Bliss}}, \bibinfo {author} {\bibfnamefont {F.}~\bibnamefont {Du}}, \bibinfo {author} {\bibfnamefont {J.}~\bibnamefont {Rao}}, \bibinfo {author} {\bibfnamefont {G.}~\bibnamefont {Song}}, \emph {et~al.},\ }\bibfield  {title} {\bibinfo {title} {{A} {R}eview of {M}agnetic {P}article {I}maging and {P}erspectives on {N}euroimaging},\ }\href@noop {} {\bibfield  {journal} {\bibinfo  {journal} {Am. J. Neuroradiol.}\ }\textbf {\bibinfo {volume} {40}},\ \bibinfo {pages} {206} (\bibinfo {year} {2019})}\BibitemShut {NoStop}%
\bibitem [{\citenamefont {Bekovi{\'c}}\ \emph {et~al.}(2023)\citenamefont {Bekovi{\'c}}, \citenamefont {Ban}, \citenamefont {Drofenik},\ and\ \citenamefont {Stergar}}]{bekovic2023magnetic}%
  \BibitemOpen
  \bibfield  {author} {\bibinfo {author} {\bibfnamefont {M.}~\bibnamefont {Bekovi{\'c}}}, \bibinfo {author} {\bibfnamefont {I.}~\bibnamefont {Ban}}, \bibinfo {author} {\bibfnamefont {M.}~\bibnamefont {Drofenik}},\ and\ \bibinfo {author} {\bibfnamefont {J.}~\bibnamefont {Stergar}},\ }\bibfield  {title} {\bibinfo {title} {{M}agnetic {N}anoparticles as {M}ediators for {M}agnetic {H}yperthermia {T}herapy {A}pplications: {A} {S}tatus {R}eview},\ }\href@noop {} {\bibfield  {journal} {\bibinfo  {journal} {Appl. Sci.}\ }\textbf {\bibinfo {volume} {13}},\ \bibinfo {pages} {9548} (\bibinfo {year} {2023})}\BibitemShut {NoStop}%
\bibitem [{\citenamefont {Fortin}\ \emph {et~al.}(2008)\citenamefont {Fortin}, \citenamefont {Gazeau},\ and\ \citenamefont {Wilhelm}}]{fortin2008intracellular}%
  \BibitemOpen
  \bibfield  {author} {\bibinfo {author} {\bibfnamefont {J.-P.}\ \bibnamefont {Fortin}}, \bibinfo {author} {\bibfnamefont {F.}~\bibnamefont {Gazeau}},\ and\ \bibinfo {author} {\bibfnamefont {C.}~\bibnamefont {Wilhelm}},\ }\bibfield  {title} {\bibinfo {title} {{I}ntracellular heating of living cells through {N}{\'e}el relaxation of magnetic nanoparticles},\ }\href@noop {} {\bibfield  {journal} {\bibinfo  {journal} {Eur. Biophys. J.}\ }\textbf {\bibinfo {volume} {37}},\ \bibinfo {pages} {223} (\bibinfo {year} {2008})}\BibitemShut {NoStop}%
\bibitem [{\citenamefont {Bogren}\ \emph {et~al.}(2015)\citenamefont {Bogren}, \citenamefont {Fornara}, \citenamefont {Ludwig}, \citenamefont {del Puerto~Morales}, \citenamefont {Steinhoff}, \citenamefont {Fougt~Hansen}, \citenamefont {Kazakova},\ and\ \citenamefont {Johansson}}]{bogren2015classification}%
  \BibitemOpen
  \bibfield  {author} {\bibinfo {author} {\bibfnamefont {S.}~\bibnamefont {Bogren}}, \bibinfo {author} {\bibfnamefont {A.}~\bibnamefont {Fornara}}, \bibinfo {author} {\bibfnamefont {F.}~\bibnamefont {Ludwig}}, \bibinfo {author} {\bibfnamefont {M.}~\bibnamefont {del Puerto~Morales}}, \bibinfo {author} {\bibfnamefont {U.}~\bibnamefont {Steinhoff}}, \bibinfo {author} {\bibfnamefont {M.}~\bibnamefont {Fougt~Hansen}}, \bibinfo {author} {\bibfnamefont {O.}~\bibnamefont {Kazakova}},\ and\ \bibinfo {author} {\bibfnamefont {C.}~\bibnamefont {Johansson}},\ }\bibfield  {title} {\bibinfo {title} {{C}lassification of {M}agnetic {N}anoparticle {S}ystems—{S}ynthesis, {S}tandardization and {A}nalysis {M}ethods in the {N}ano{M}ag {P}roject},\ }\href@noop {} {\bibfield  {journal} {\bibinfo  {journal} {Int. J. Mol. Sci.}\ }\textbf {\bibinfo {volume} {16}},\ \bibinfo {pages} {20308} (\bibinfo {year} {2015})}\BibitemShut {NoStop}%
\bibitem [{\citenamefont {Chalifour}\ \emph {et~al.}(2021)\citenamefont {Chalifour}, \citenamefont {Davidson}, \citenamefont {Anderson}, \citenamefont {Crawford},\ and\ \citenamefont {Livesey}}]{chalifour2021magnetic}%
  \BibitemOpen
  \bibfield  {author} {\bibinfo {author} {\bibfnamefont {A.~R.}\ \bibnamefont {Chalifour}}, \bibinfo {author} {\bibfnamefont {J.~C.}\ \bibnamefont {Davidson}}, \bibinfo {author} {\bibfnamefont {N.~R.}\ \bibnamefont {Anderson}}, \bibinfo {author} {\bibfnamefont {T.~M.}\ \bibnamefont {Crawford}},\ and\ \bibinfo {author} {\bibfnamefont {K.~L.}\ \bibnamefont {Livesey}},\ }\bibfield  {title} {\bibinfo {title} {{M}agnetic relaxation time for an ensemble of nanoparticles with randomly aligned easy axes: {A} simple expression},\ }\href@noop {} {\bibfield  {journal} {\bibinfo  {journal} {Phys. Rev. B}\ }\textbf {\bibinfo {volume} {104}},\ \bibinfo {pages} {094433} (\bibinfo {year} {2021})}\BibitemShut {NoStop}%
\bibitem [{\citenamefont {Yoshida}\ and\ \citenamefont {Enpuku}(2009)}]{yoshida2009simulation}%
  \BibitemOpen
  \bibfield  {author} {\bibinfo {author} {\bibfnamefont {T.}~\bibnamefont {Yoshida}}\ and\ \bibinfo {author} {\bibfnamefont {K.}~\bibnamefont {Enpuku}},\ }\bibfield  {title} {\bibinfo {title} {{S}imulation and {Q}uantitative {C}larification of {AC} {S}usceptibility of {M}agnetic {F}luid in {N}onlinear {B}rownian {R}elaxation {R}egion},\ }\href@noop {} {\bibfield  {journal} {\bibinfo  {journal} {Jpn. J. Appl. Phys.}\ }\textbf {\bibinfo {volume} {48}},\ \bibinfo {pages} {127002} (\bibinfo {year} {2009})}\BibitemShut {NoStop}%
\bibitem [{\citenamefont {Martsenyuk}\ \emph {et~al.}(1974)\citenamefont {Martsenyuk}, \citenamefont {Raikher},\ and\ \citenamefont {Shliomis}}]{martsenyuk1974kinetics}%
  \BibitemOpen
  \bibfield  {author} {\bibinfo {author} {\bibfnamefont {M.}~\bibnamefont {Martsenyuk}}, \bibinfo {author} {\bibfnamefont {Y.~L.}\ \bibnamefont {Raikher}},\ and\ \bibinfo {author} {\bibfnamefont {M.}~\bibnamefont {Shliomis}},\ }\bibfield  {title} {\bibinfo {title} {{O}n the kinetics of magnetization of suspension of ferromagnetic particles},\ }\href@noop {} {\bibfield  {journal} {\bibinfo  {journal} {Sov. Phys. JETP}\ }\textbf {\bibinfo {volume} {38}},\ \bibinfo {pages} {413} (\bibinfo {year} {1974})}\BibitemShut {NoStop}%
\bibitem [{\citenamefont {Dieckhoff}\ \emph {et~al.}(2016)\citenamefont {Dieckhoff}, \citenamefont {Eberbeck}, \citenamefont {Schilling},\ and\ \citenamefont {Ludwig}}]{dieckhoff2016magnetic}%
  \BibitemOpen
  \bibfield  {author} {\bibinfo {author} {\bibfnamefont {J.}~\bibnamefont {Dieckhoff}}, \bibinfo {author} {\bibfnamefont {D.}~\bibnamefont {Eberbeck}}, \bibinfo {author} {\bibfnamefont {M.}~\bibnamefont {Schilling}},\ and\ \bibinfo {author} {\bibfnamefont {F.}~\bibnamefont {Ludwig}},\ }\bibfield  {title} {\bibinfo {title} {Magnetic-field dependence of {Brownian and N{\'e}el} relaxation times},\ }\href@noop {} {\bibfield  {journal} {\bibinfo  {journal} {J. Appl. Phys.}\ }\textbf {\bibinfo {volume} {119}},\ \bibinfo {pages} {043903} (\bibinfo {year} {2016})}\BibitemShut {NoStop}%
\bibitem [{\citenamefont {Brown~Jr}(1963)}]{brown1963thermal}%
  \BibitemOpen
  \bibfield  {author} {\bibinfo {author} {\bibfnamefont {W.~F.}\ \bibnamefont {Brown~Jr}},\ }\bibfield  {title} {\bibinfo {title} {{T}hermal {F}luctuations of a {S}ingle-{D}omain {P}article},\ }\href@noop {} {\bibfield  {journal} {\bibinfo  {journal} {Phys. Rev.}\ }\textbf {\bibinfo {volume} {130}},\ \bibinfo {pages} {1677} (\bibinfo {year} {1963})}\BibitemShut {NoStop}%
\bibitem [{\citenamefont {Deissler}\ \emph {et~al.}(2014)\citenamefont {Deissler}, \citenamefont {Wu},\ and\ \citenamefont {Martens}}]{deissler2014dependence}%
  \BibitemOpen
  \bibfield  {author} {\bibinfo {author} {\bibfnamefont {R.~J.}\ \bibnamefont {Deissler}}, \bibinfo {author} {\bibfnamefont {Y.}~\bibnamefont {Wu}},\ and\ \bibinfo {author} {\bibfnamefont {M.~A.}\ \bibnamefont {Martens}},\ }\bibfield  {title} {\bibinfo {title} {{D}ependence of {B}rownian and {N}{\'e}el relaxation times on magnetic field strength},\ }\href@noop {} {\bibfield  {journal} {\bibinfo  {journal} {Med. Phys.}\ }\textbf {\bibinfo {volume} {41}},\ \bibinfo {pages} {012301} (\bibinfo {year} {2014})}\BibitemShut {NoStop}%
\bibitem [{\citenamefont {Ota}\ \emph {et~al.}(2016)\citenamefont {Ota}, \citenamefont {Kitaguchi}, \citenamefont {Takeda}, \citenamefont {Yamada},\ and\ \citenamefont {Takemura}}]{ota2016rotation}%
  \BibitemOpen
  \bibfield  {author} {\bibinfo {author} {\bibfnamefont {S.}~\bibnamefont {Ota}}, \bibinfo {author} {\bibfnamefont {R.}~\bibnamefont {Kitaguchi}}, \bibinfo {author} {\bibfnamefont {R.}~\bibnamefont {Takeda}}, \bibinfo {author} {\bibfnamefont {T.}~\bibnamefont {Yamada}},\ and\ \bibinfo {author} {\bibfnamefont {Y.}~\bibnamefont {Takemura}},\ }\bibfield  {title} {\bibinfo {title} {{R}otation of {M}agnetization {D}erived from {B}rownian {R}elaxation in {M}agnetic {F}luids of {D}ifferent {V}iscosity {E}valuated by {D}ynamic {H}ysteresis {M}easurements over a {W}ide {F}requency {R}ange},\ }\href@noop {} {\bibfield  {journal} {\bibinfo  {journal} {Nanomaterials}\ }\textbf {\bibinfo {volume} {6}},\ \bibinfo {pages} {170} (\bibinfo {year} {2016})}\BibitemShut {NoStop}%
\bibitem [{\citenamefont {Lemons}\ and\ \citenamefont {Gythiel}(1997)}]{lemons1997paul}%
  \BibitemOpen
  \bibfield  {author} {\bibinfo {author} {\bibfnamefont {D.~S.}\ \bibnamefont {Lemons}}\ and\ \bibinfo {author} {\bibfnamefont {A.}~\bibnamefont {Gythiel}},\ }\bibfield  {title} {\bibinfo {title} {Paul {L}angevin’s 1908 paper “{O}n the {T}heory of {B}rownian {M}otion” [“{S}ur la théorie du mouvement brownien,” {C}. {R}. {A}cad. {S}ci. ({P}aris) 146, 530–533 (1908)]},\ }\href@noop {} {\bibfield  {journal} {\bibinfo  {journal} {Am. J. Phys.}\ }\textbf {\bibinfo {volume} {65}},\ \bibinfo {pages} {1079} (\bibinfo {year} {1997})}\BibitemShut {NoStop}%
\bibitem [{\citenamefont {Berkov}\ \emph {et~al.}(2006)\citenamefont {Berkov}, \citenamefont {Gorn}, \citenamefont {Schmitz},\ and\ \citenamefont {Stock}}]{berkov2006langevin}%
  \BibitemOpen
  \bibfield  {author} {\bibinfo {author} {\bibfnamefont {D.}~\bibnamefont {Berkov}}, \bibinfo {author} {\bibfnamefont {N.}~\bibnamefont {Gorn}}, \bibinfo {author} {\bibfnamefont {R.}~\bibnamefont {Schmitz}},\ and\ \bibinfo {author} {\bibfnamefont {D.}~\bibnamefont {Stock}},\ }\bibfield  {title} {\bibinfo {title} {{L}angevin dynamic simulations of fast remagnetization processes in ferrofluids with internal magnetic degrees of freedom},\ }\href@noop {} {\bibfield  {journal} {\bibinfo  {journal} {J. Phys.-Condens. Mat.}\ }\textbf {\bibinfo {volume} {18}},\ \bibinfo {pages} {S2595} (\bibinfo {year} {2006})}\BibitemShut {NoStop}%
\bibitem [{\citenamefont {Kittel}(1949)}]{kittel1949physical}%
  \BibitemOpen
  \bibfield  {author} {\bibinfo {author} {\bibfnamefont {C.}~\bibnamefont {Kittel}},\ }\bibfield  {title} {\bibinfo {title} {{P}hysical {T}heory of {F}erromagnetic {D}omains},\ }\href@noop {} {\bibfield  {journal} {\bibinfo  {journal} {Rev. Mod. Phys.}\ }\textbf {\bibinfo {volume} {21}},\ \bibinfo {pages} {541} (\bibinfo {year} {1949})}\BibitemShut {NoStop}%
\bibitem [{\citenamefont {Reichel}\ \emph {et~al.}(2017)\citenamefont {Reichel}, \citenamefont {Kov{\'a}cs}, \citenamefont {Kumari}, \citenamefont {Bereczk-Tompa}, \citenamefont {Schneck}, \citenamefont {Diehle}, \citenamefont {P{\'o}sfai}, \citenamefont {Hirt}, \citenamefont {Duchamp}, \citenamefont {Dunin-Borkowski} \emph {et~al.}}]{reichel2017single}%
  \BibitemOpen
  \bibfield  {author} {\bibinfo {author} {\bibfnamefont {V.}~\bibnamefont {Reichel}}, \bibinfo {author} {\bibfnamefont {A.}~\bibnamefont {Kov{\'a}cs}}, \bibinfo {author} {\bibfnamefont {M.}~\bibnamefont {Kumari}}, \bibinfo {author} {\bibfnamefont {{\'E}.}~\bibnamefont {Bereczk-Tompa}}, \bibinfo {author} {\bibfnamefont {E.}~\bibnamefont {Schneck}}, \bibinfo {author} {\bibfnamefont {P.}~\bibnamefont {Diehle}}, \bibinfo {author} {\bibfnamefont {M.}~\bibnamefont {P{\'o}sfai}}, \bibinfo {author} {\bibfnamefont {A.~M.}\ \bibnamefont {Hirt}}, \bibinfo {author} {\bibfnamefont {M.}~\bibnamefont {Duchamp}}, \bibinfo {author} {\bibfnamefont {R.~E.}\ \bibnamefont {Dunin-Borkowski}}, \emph {et~al.},\ }\bibfield  {title} {\bibinfo {title} {{S}ingle crystalline superstructured stable single domain magnetite nanoparticles},\ }\href@noop {} {\bibfield  {journal} {\bibinfo  {journal} {Sci. Rep.}\ }\textbf {\bibinfo {volume} {7}},\ \bibinfo {pages} {45484} (\bibinfo {year} {2017})}\BibitemShut {NoStop}%
\bibitem [{\citenamefont {Debye}(1929)}]{debye1929polar}%
  \BibitemOpen
  \bibfield  {author} {\bibinfo {author} {\bibfnamefont {P.}~\bibnamefont {Debye}},\ }\href {https://books.google.com/books?id=2WuUwgEACAAJ} {\emph {\bibinfo {title} {Polar Molecules}}}\ (\bibinfo  {publisher} {Chemical Catalog Co},\ \bibinfo {address} {New York},\ \bibinfo {year} {1929})\BibitemShut {NoStop}%
\bibitem [{\citenamefont {Anderson}\ \emph {et~al.}(2021)\citenamefont {Anderson}, \citenamefont {Davidson}, \citenamefont {Louie}, \citenamefont {Serantes},\ and\ \citenamefont {Livesey}}]{anderson2021simulating}%
  \BibitemOpen
  \bibfield  {author} {\bibinfo {author} {\bibfnamefont {N.~R.}\ \bibnamefont {Anderson}}, \bibinfo {author} {\bibfnamefont {J.}~\bibnamefont {Davidson}}, \bibinfo {author} {\bibfnamefont {D.~R.}\ \bibnamefont {Louie}}, \bibinfo {author} {\bibfnamefont {D.}~\bibnamefont {Serantes}},\ and\ \bibinfo {author} {\bibfnamefont {K.~L.}\ \bibnamefont {Livesey}},\ }\bibfield  {title} {\bibinfo {title} {{S}imulating the {S}elf-{A}ssembly and {H}ysteresis {L}oops of {F}erromagnetic {N}anoparticles with {S}ticking of {L}igands},\ }\href@noop {} {\bibfield  {journal} {\bibinfo  {journal} {Nanomat.}\ }\textbf {\bibinfo {volume} {11}},\ \bibinfo {pages} {2870} (\bibinfo {year} {2021})}\BibitemShut {NoStop}%
\bibitem [{\citenamefont {Wang}\ \emph {et~al.}(2002)\citenamefont {Wang}, \citenamefont {Holm},\ and\ \citenamefont {M{\"u}ller}}]{wang2002molecular}%
  \BibitemOpen
  \bibfield  {author} {\bibinfo {author} {\bibfnamefont {Z.}~\bibnamefont {Wang}}, \bibinfo {author} {\bibfnamefont {C.}~\bibnamefont {Holm}},\ and\ \bibinfo {author} {\bibfnamefont {H.~W.}\ \bibnamefont {M{\"u}ller}},\ }\bibfield  {title} {\bibinfo {title} {{M}olecular dynamics study on the equilibrium magnetization properties and structure of ferrofluids},\ }\href@noop {} {\bibfield  {journal} {\bibinfo  {journal} {Phys. Rev. E}\ }\textbf {\bibinfo {volume} {66}},\ \bibinfo {pages} {021405} (\bibinfo {year} {2002})}\BibitemShut {NoStop}%
\bibitem [{\citenamefont {Weisstein}(2002)}]{weisstein2002sphere}%
  \BibitemOpen
  \bibfield  {author} {\bibinfo {author} {\bibfnamefont {E.~W.}\ \bibnamefont {Weisstein}},\ }\bibfield  {title} {\bibinfo {title} {{S}phere {P}oint {P}icking},\ }\href@noop {} {\bibfield  {journal} {\bibinfo  {journal} {https://mathworld. wolfram. com/}\ } (\bibinfo {year} {2002})}\BibitemShut {NoStop}%
\bibitem [{\citenamefont {Jensen}\ and\ \citenamefont {Finlayson}(1980)}]{jensen1980solution}%
  \BibitemOpen
  \bibfield  {author} {\bibinfo {author} {\bibfnamefont {O.~K.}\ \bibnamefont {Jensen}}\ and\ \bibinfo {author} {\bibfnamefont {B.~A.}\ \bibnamefont {Finlayson}},\ }\bibfield  {title} {\bibinfo {title} {{S}olution of the transport equations using a moving coordinate system},\ }\href@noop {} {\bibfield  {journal} {\bibinfo  {journal} {Adv. Water Resour.}\ }\textbf {\bibinfo {volume} {3}},\ \bibinfo {pages} {9} (\bibinfo {year} {1980})}\BibitemShut {NoStop}%
\bibitem [{\citenamefont {Saville}\ \emph {et~al.}(2014)\citenamefont {Saville}, \citenamefont {Qi}, \citenamefont {Baker}, \citenamefont {Stone}, \citenamefont {Camley}, \citenamefont {Livesey}, \citenamefont {Ye}, \citenamefont {Crawford},\ and\ \citenamefont {Mefford}}]{saville2014formation}%
  \BibitemOpen
  \bibfield  {author} {\bibinfo {author} {\bibfnamefont {S.~L.}\ \bibnamefont {Saville}}, \bibinfo {author} {\bibfnamefont {B.}~\bibnamefont {Qi}}, \bibinfo {author} {\bibfnamefont {J.}~\bibnamefont {Baker}}, \bibinfo {author} {\bibfnamefont {R.}~\bibnamefont {Stone}}, \bibinfo {author} {\bibfnamefont {R.~E.}\ \bibnamefont {Camley}}, \bibinfo {author} {\bibfnamefont {K.~L.}\ \bibnamefont {Livesey}}, \bibinfo {author} {\bibfnamefont {L.}~\bibnamefont {Ye}}, \bibinfo {author} {\bibfnamefont {T.~M.}\ \bibnamefont {Crawford}},\ and\ \bibinfo {author} {\bibfnamefont {O.~T.}\ \bibnamefont {Mefford}},\ }\bibfield  {title} {\bibinfo {title} {{T}he formation of linear aggregates in magnetic hyperthermia: {I}mplications on specific absorption rate and magnetic anisotropy},\ }\href@noop {} {\bibfield  {journal} {\bibinfo  {journal} {J. Colloid. Interf. Sci.}\ }\textbf {\bibinfo {volume} {424}},\ \bibinfo {pages} {141} (\bibinfo {year} {2014})}\BibitemShut {NoStop}%
\bibitem [{\citenamefont {Mamiya}\ \emph {et~al.}(2020)\citenamefont {Mamiya}, \citenamefont {Fukumoto}, \citenamefont {Cuya~Huaman}, \citenamefont {Suzuki}, \citenamefont {Miyamura},\ and\ \citenamefont {Balachandran}}]{mamiya2020estimation}%
  \BibitemOpen
  \bibfield  {author} {\bibinfo {author} {\bibfnamefont {H.}~\bibnamefont {Mamiya}}, \bibinfo {author} {\bibfnamefont {H.}~\bibnamefont {Fukumoto}}, \bibinfo {author} {\bibfnamefont {J.~L.}\ \bibnamefont {Cuya~Huaman}}, \bibinfo {author} {\bibfnamefont {K.}~\bibnamefont {Suzuki}}, \bibinfo {author} {\bibfnamefont {H.}~\bibnamefont {Miyamura}},\ and\ \bibinfo {author} {\bibfnamefont {J.}~\bibnamefont {Balachandran}},\ }\bibfield  {title} {\bibinfo {title} {{E}stimation of {M}agnetic {A}nisotropy of {I}ndividual {M}agnetite {N}anoparticles for {M}agnetic {H}yperthermia},\ }\href@noop {} {\bibfield  {journal} {\bibinfo  {journal} {ACS nano}\ }\textbf {\bibinfo {volume} {14}},\ \bibinfo {pages} {8421} (\bibinfo {year} {2020})}\BibitemShut {NoStop}%
\bibitem [{\citenamefont {Fannin}\ and\ \citenamefont {Charles}(1994)}]{fannin1994calculation}%
  \BibitemOpen
  \bibfield  {author} {\bibinfo {author} {\bibfnamefont {P.}~\bibnamefont {Fannin}}\ and\ \bibinfo {author} {\bibfnamefont {S.}~\bibnamefont {Charles}},\ }\bibfield  {title} {\bibinfo {title} {{O}n the calculation of the {N}eel relaxation time in uniaxial single-domain ferromagnetic particles},\ }\href@noop {} {\bibfield  {journal} {\bibinfo  {journal} {J. Phys. D Appl. Phys.}\ }\textbf {\bibinfo {volume} {27}},\ \bibinfo {pages} {185} (\bibinfo {year} {1994})}\BibitemShut {NoStop}%
\bibitem [{\citenamefont {Yoon}(2011)}]{yoon2011determination}%
  \BibitemOpen
  \bibfield  {author} {\bibinfo {author} {\bibfnamefont {S.}~\bibnamefont {Yoon}},\ }\bibfield  {title} {\bibinfo {title} {{D}etermination of the {T}emperature {D}ependence of the {M}agnetic {A}nisotropy {C}onstant in {M}agnetite {N}anoparticles},\ }\href@noop {} {\bibfield  {journal} {\bibinfo  {journal} {J. Korean Phys. Soc.}\ }\textbf {\bibinfo {volume} {59}},\ \bibinfo {pages} {3069} (\bibinfo {year} {2011})}\BibitemShut {NoStop}%
\bibitem [{\citenamefont {Zhao}\ \emph {et~al.}(2020)\citenamefont {Zhao}, \citenamefont {Garraud}, \citenamefont {Arnold},\ and\ \citenamefont {Rinaldi-Ramos}}]{zhao2020effects}%
  \BibitemOpen
  \bibfield  {author} {\bibinfo {author} {\bibfnamefont {Z.}~\bibnamefont {Zhao}}, \bibinfo {author} {\bibfnamefont {N.}~\bibnamefont {Garraud}}, \bibinfo {author} {\bibfnamefont {D.~P.}\ \bibnamefont {Arnold}},\ and\ \bibinfo {author} {\bibfnamefont {C.}~\bibnamefont {Rinaldi-Ramos}},\ }\bibfield  {title} {\bibinfo {title} {Effects of particle diameter and magnetocrystalline anisotropy on magnetic relaxation and magnetic particle imaging performance of magnetic nanoparticles},\ }\href@noop {} {\bibfield  {journal} {\bibinfo  {journal} {Phys. Med. Biol.}\ }\textbf {\bibinfo {volume} {65}},\ \bibinfo {pages} {025014} (\bibinfo {year} {2020})}\BibitemShut {NoStop}%
\bibitem [{\citenamefont {Engelmann}\ \emph {et~al.}(2019)\citenamefont {Engelmann}, \citenamefont {Shasha}, \citenamefont {Teeman}, \citenamefont {Slabu},\ and\ \citenamefont {Krishnan}}]{engelmann2019predicting}%
  \BibitemOpen
  \bibfield  {author} {\bibinfo {author} {\bibfnamefont {U.~M.}\ \bibnamefont {Engelmann}}, \bibinfo {author} {\bibfnamefont {C.}~\bibnamefont {Shasha}}, \bibinfo {author} {\bibfnamefont {E.}~\bibnamefont {Teeman}}, \bibinfo {author} {\bibfnamefont {I.}~\bibnamefont {Slabu}},\ and\ \bibinfo {author} {\bibfnamefont {K.~M.}\ \bibnamefont {Krishnan}},\ }\bibfield  {title} {\bibinfo {title} {{P}redicting size-dependent heating efficiency of magnetic nanoparticles from experiment and stochastic {N}{\'e}el-{B}rown {L}angevin simulation},\ }\href@noop {} {\bibfield  {journal} {\bibinfo  {journal} {J. Magn. Magn. Mater.}\ }\textbf {\bibinfo {volume} {471}},\ \bibinfo {pages} {450} (\bibinfo {year} {2019})}\BibitemShut {NoStop}%
\bibitem [{\citenamefont {Dormann}\ \emph {et~al.}(1997)\citenamefont {Dormann}, \citenamefont {Fiorani},\ and\ \citenamefont {Tronc}}]{dormann1997magnetic}%
  \BibitemOpen
  \bibfield  {author} {\bibinfo {author} {\bibfnamefont {J.-L.}\ \bibnamefont {Dormann}}, \bibinfo {author} {\bibfnamefont {D.}~\bibnamefont {Fiorani}},\ and\ \bibinfo {author} {\bibfnamefont {E.}~\bibnamefont {Tronc}},\ }\bibfield  {title} {\bibinfo {title} {Magnetic relaxation in fine-particle systems},\ }\href@noop {} {\bibfield  {journal} {\bibinfo  {journal} {Adv. Chem. Phys.}\ }\textbf {\bibinfo {volume} {98}},\ \bibinfo {pages} {283} (\bibinfo {year} {1997})}\BibitemShut {NoStop}%
\bibitem [{\citenamefont {Ilg}\ and\ \citenamefont {Kr{\"o}ger}(2020)}]{ilg2020dynamics}%
  \BibitemOpen
  \bibfield  {author} {\bibinfo {author} {\bibfnamefont {P.}~\bibnamefont {Ilg}}\ and\ \bibinfo {author} {\bibfnamefont {M.}~\bibnamefont {Kr{\"o}ger}},\ }\bibfield  {title} {\bibinfo {title} {{D}ynamics of interacting magnetic nanoparticles: {E}ffective behavior from competition between {B}rownian and {N}{\'e}el relaxation},\ }\href@noop {} {\bibfield  {journal} {\bibinfo  {journal} {Phys. Chem. Chem. Phys.}\ }\textbf {\bibinfo {volume} {22}},\ \bibinfo {pages} {22244} (\bibinfo {year} {2020})}\BibitemShut {NoStop}%
\bibitem [{\citenamefont {Nader}\ \emph {et~al.}(2019)\citenamefont {Nader}, \citenamefont {Skinner}, \citenamefont {Romana}, \citenamefont {Fort}, \citenamefont {Lemonne}, \citenamefont {Guillot}, \citenamefont {Gauthier}, \citenamefont {Antoine-Jonville}, \citenamefont {Renoux}, \citenamefont {Hardy-Dessources} \emph {et~al.}}]{nader2019blood}%
  \BibitemOpen
  \bibfield  {author} {\bibinfo {author} {\bibfnamefont {E.}~\bibnamefont {Nader}}, \bibinfo {author} {\bibfnamefont {S.}~\bibnamefont {Skinner}}, \bibinfo {author} {\bibfnamefont {M.}~\bibnamefont {Romana}}, \bibinfo {author} {\bibfnamefont {R.}~\bibnamefont {Fort}}, \bibinfo {author} {\bibfnamefont {N.}~\bibnamefont {Lemonne}}, \bibinfo {author} {\bibfnamefont {N.}~\bibnamefont {Guillot}}, \bibinfo {author} {\bibfnamefont {A.}~\bibnamefont {Gauthier}}, \bibinfo {author} {\bibfnamefont {S.}~\bibnamefont {Antoine-Jonville}}, \bibinfo {author} {\bibfnamefont {C.}~\bibnamefont {Renoux}}, \bibinfo {author} {\bibfnamefont {M.-D.}\ \bibnamefont {Hardy-Dessources}}, \emph {et~al.},\ }\bibfield  {title} {\bibinfo {title} {Blood rheology: key parameters, impact on blood flow, role in sickle cell disease and effects of exercise},\ }\href@noop {} {\bibfield  {journal} {\bibinfo  {journal} {Front. Physiol.}\ }\textbf {\bibinfo {volume} {10}},\ \bibinfo {pages} {1329} (\bibinfo {year} {2019})}\BibitemShut {NoStop}%
\bibitem [{\citenamefont {Butcher}(1963)}]{butcher1963coefficients}%
  \BibitemOpen
  \bibfield  {author} {\bibinfo {author} {\bibfnamefont {J.~C.}\ \bibnamefont {Butcher}},\ }\bibfield  {title} {\bibinfo {title} {{C}oefficients for the study of {R}unge-{K}utta integration processes},\ }\href@noop {} {\bibfield  {journal} {\bibinfo  {journal} {J. Aust. Math. Soc.}\ }\textbf {\bibinfo {volume} {3}},\ \bibinfo {pages} {185} (\bibinfo {year} {1963})}\BibitemShut {NoStop}%
\bibitem [{\citenamefont {Brown}(1962)}]{brown1962magnetostatic}%
  \BibitemOpen
  \bibfield  {author} {\bibinfo {author} {\bibfnamefont {W.~F.}\ \bibnamefont {Brown}},\ }\href@noop {} {\emph {\bibinfo {title} {Magnetostatic principles in {F}erromagnetism}}},\ Vol.~\bibinfo {volume} {1}\ (\bibinfo  {publisher} {North-Holland Publishing Company},\ \bibinfo {year} {1962})\BibitemShut {NoStop}%
\bibitem [{\citenamefont {Tannous}\ and\ \citenamefont {Gieraltowski}(2008)}]{tannous2008stoner}%
  \BibitemOpen
  \bibfield  {author} {\bibinfo {author} {\bibfnamefont {C.}~\bibnamefont {Tannous}}\ and\ \bibinfo {author} {\bibfnamefont {J.}~\bibnamefont {Gieraltowski}},\ }\bibfield  {title} {\bibinfo {title} {The {Stoner--Wohlfarth} model of ferromagnetism},\ }\href@noop {} {\bibfield  {journal} {\bibinfo  {journal} {Eur. J. Phys.}\ }\textbf {\bibinfo {volume} {29}},\ \bibinfo {pages} {475} (\bibinfo {year} {2008})}\BibitemShut {NoStop}%
\bibitem [{\citenamefont {Chantrell}\ \emph {et~al.}(2000)\citenamefont {Chantrell}, \citenamefont {Walmsley}, \citenamefont {Gore},\ and\ \citenamefont {Maylin}}]{chantrell2000calculations}%
  \BibitemOpen
  \bibfield  {author} {\bibinfo {author} {\bibfnamefont {R.}~\bibnamefont {Chantrell}}, \bibinfo {author} {\bibfnamefont {N.}~\bibnamefont {Walmsley}}, \bibinfo {author} {\bibfnamefont {J.}~\bibnamefont {Gore}},\ and\ \bibinfo {author} {\bibfnamefont {M.}~\bibnamefont {Maylin}},\ }\bibfield  {title} {\bibinfo {title} {{C}alculations of the susceptibility of interacting superparamagnetic particles},\ }\href@noop {} {\bibfield  {journal} {\bibinfo  {journal} {Phys. Rev. B}\ }\textbf {\bibinfo {volume} {63}},\ \bibinfo {pages} {024410} (\bibinfo {year} {2000})}\BibitemShut {NoStop}%
\bibitem [{\citenamefont {Nunes}\ \emph {et~al.}(2005)\citenamefont {Nunes}, \citenamefont {Socolovsky}, \citenamefont {Denardin}, \citenamefont {Cebollada}, \citenamefont {Brandl},\ and\ \citenamefont {Knobel}}]{nunes2005role}%
  \BibitemOpen
  \bibfield  {author} {\bibinfo {author} {\bibfnamefont {W.}~\bibnamefont {Nunes}}, \bibinfo {author} {\bibfnamefont {L.}~\bibnamefont {Socolovsky}}, \bibinfo {author} {\bibfnamefont {J.}~\bibnamefont {Denardin}}, \bibinfo {author} {\bibfnamefont {F.}~\bibnamefont {Cebollada}}, \bibinfo {author} {\bibfnamefont {A.~L.}\ \bibnamefont {Brandl}},\ and\ \bibinfo {author} {\bibfnamefont {M.}~\bibnamefont {Knobel}},\ }\bibfield  {title} {\bibinfo {title} {{R}ole of magnetic interparticle coupling on the field dependence of the superparamagnetic relaxation time},\ }\href@noop {} {\bibfield  {journal} {\bibinfo  {journal} {Phys. Rev. B}\ }\textbf {\bibinfo {volume} {72}},\ \bibinfo {pages} {212413} (\bibinfo {year} {2005})}\BibitemShut {NoStop}%
\bibitem [{\citenamefont {Kim}\ \emph {et~al.}(2015)\citenamefont {Kim}, \citenamefont {Mangal},\ and\ \citenamefont {Archer}}]{kim2015relaxation}%
  \BibitemOpen
  \bibfield  {author} {\bibinfo {author} {\bibfnamefont {S.~A.}\ \bibnamefont {Kim}}, \bibinfo {author} {\bibfnamefont {R.}~\bibnamefont {Mangal}},\ and\ \bibinfo {author} {\bibfnamefont {L.~A.}\ \bibnamefont {Archer}},\ }\bibfield  {title} {\bibinfo {title} {{R}elaxation {D}ynamics of {N}anoparticle-{T}ethered {P}olymer {C}hains},\ }\href@noop {} {\bibfield  {journal} {\bibinfo  {journal} {Macromolecules}\ }\textbf {\bibinfo {volume} {48}},\ \bibinfo {pages} {6280} (\bibinfo {year} {2015})}\BibitemShut {NoStop}%
\bibitem [{\citenamefont {Livesey}\ \emph {et~al.}(2018)\citenamefont {Livesey}, \citenamefont {Ruta}, \citenamefont {Anderson}, \citenamefont {Baldomir}, \citenamefont {Chantrell},\ and\ \citenamefont {Serantes}}]{livesey2018beyond}%
  \BibitemOpen
  \bibfield  {author} {\bibinfo {author} {\bibfnamefont {K.~L.}\ \bibnamefont {Livesey}}, \bibinfo {author} {\bibfnamefont {S.}~\bibnamefont {Ruta}}, \bibinfo {author} {\bibfnamefont {N.}~\bibnamefont {Anderson}}, \bibinfo {author} {\bibfnamefont {D.}~\bibnamefont {Baldomir}}, \bibinfo {author} {\bibfnamefont {R.~W.}\ \bibnamefont {Chantrell}},\ and\ \bibinfo {author} {\bibfnamefont {D.}~\bibnamefont {Serantes}},\ }\bibfield  {title} {\bibinfo {title} {{B}eyond the blocking model to fit nanoparticle {ZFC}/{FC} magnetisation curves},\ }\href@noop {} {\bibfield  {journal} {\bibinfo  {journal} {Sci. Rep.}\ }\textbf {\bibinfo {volume} {8}},\ \bibinfo {pages} {1} (\bibinfo {year} {2018})}\BibitemShut {NoStop}%
\bibitem [{\citenamefont {Respaud}\ \emph {et~al.}(1998)\citenamefont {Respaud}, \citenamefont {Broto}, \citenamefont {Rakoto}, \citenamefont {Fert}, \citenamefont {Thomas}, \citenamefont {Barbara}, \citenamefont {Verelst}, \citenamefont {Snoeck}, \citenamefont {Lecante}, \citenamefont {Mosset} \emph {et~al.}}]{respaud1998surface}%
  \BibitemOpen
  \bibfield  {author} {\bibinfo {author} {\bibfnamefont {M.}~\bibnamefont {Respaud}}, \bibinfo {author} {\bibfnamefont {J.}~\bibnamefont {Broto}}, \bibinfo {author} {\bibfnamefont {H.}~\bibnamefont {Rakoto}}, \bibinfo {author} {\bibfnamefont {A.}~\bibnamefont {Fert}}, \bibinfo {author} {\bibfnamefont {L.}~\bibnamefont {Thomas}}, \bibinfo {author} {\bibfnamefont {B.}~\bibnamefont {Barbara}}, \bibinfo {author} {\bibfnamefont {M.}~\bibnamefont {Verelst}}, \bibinfo {author} {\bibfnamefont {E.}~\bibnamefont {Snoeck}}, \bibinfo {author} {\bibfnamefont {P.}~\bibnamefont {Lecante}}, \bibinfo {author} {\bibfnamefont {A.}~\bibnamefont {Mosset}}, \emph {et~al.},\ }\bibfield  {title} {\bibinfo {title} {{S}urface effects on the magnetic properties of ultrafine cobalt particles},\ }\href@noop {} {\bibfield  {journal} {\bibinfo  {journal} {Phys. Rev. B}\ }\textbf {\bibinfo {volume} {57}},\ \bibinfo {pages} {2925} (\bibinfo {year} {1998})}\BibitemShut {NoStop}%
\end{thebibliography}%

\end{document}